\documentclass[Afour,sageh,times]{sagej}

\usepackage{moreverb,url}

\usepackage[colorlinks,bookmarksopen,bookmarksnumbered,citecolor=red,urlcolor=red]{hyperref}
\usepackage{xcolor}

\usepackage{listings}
\newcommand\BibTeX{{\rmfamily B\kern-.05em \textsc{i\kern-.025em b}\kern-.08em
T\kern-.1667em\lower.7ex\hbox{E}\kern-.125emX}}

\def\volumeyear{2022}

\usepackage{multirow}

\usepackage{xspace}
\newcommand{\codename}[1]{\textsc{#1}}
\newcommand{\parthenon}{\codename{Parthenon}\xspace}
\newcommand{\enzo}{\codename{Enzo}\xspace}
\newcommand{\gamer}{\codename{gamer-2}\xspace}
\newcommand{\paramesh}{\codename{PARAMESH}\xspace}
\newcommand{\rage}{\codename{xRAGE}\xspace}
\newcommand{\ramses}{\codename{RAMSES}\xspace}
\newcommand{\pluto}{\codename{Pluto}\xspace}
\newcommand{\athena}{\codename{Athena++}\xspace}
\newcommand{\charm}{\codename{Charm++}\xspace}
\newcommand{\athenapk}{\codename{AthenaPK}\xspace}
\newcommand{\kathena}{\codename{K-Athena}\xspace}
\newcommand{\phydro}{\codename{Parthenon-hydro}\xspace}
\newcommand{\phoebus}{\codename{Phoebus}\xspace}
\newcommand{\singularityeos}{\codename{Singularity-EOS}\xspace}
\newcommand{\singularityopac}{\codename{Singularity-Opac}\xspace}
\newcommand{\riot}{\codename{RIOT}\xspace}
\newcommand{\amrex}{\codename{AMReX}\xspace}
\newcommand{\uintah}{\codename{Uintah}\xspace}
\newcommand{\kokkos}{\codename{Kokkos}\xspace}
\newcommand{\mpi}{\codename{MPI}\xspace}
\newcommand{\hdf}{\codename{HDF5}\xspace}
\newcommand{\ctest}{\codename{ctest}\xspace}
\newcommand{\github}{\codename{GitHub}\xspace}
\newcommand{\cpplint}{\codename{cpplint}\xspace}
\newcommand{\clangformat}{\codename{clang-format}\xspace}
\newcommand{\catch}{\codename{Catch2}\xspace}
\newcommand{\nvcc}{\codename{nvcc}\xspace}
\newcommand{\gcc}{\codename{G++}\xspace}
\newcommand{\hipcc}{\codename{hipcc}\xspace}
\newcommand{\docker}{\codename{docker}\xspace}
\newcommand{\json}{\codename{json}\xspace}
\newcommand{\python}{\codename{Python}\xspace}
\newcommand{\yt}{\codename{yt}\xspace}
\newcommand{\paraview}{\codename{ParaView}\xspace}
\newcommand{\visit}{\codename{VisIt}\xspace}
\newcommand{\flash}{\codename{Flash}\xspace}
\newcommand{\quinoa}{\codename{Quinoa}\xspace}

\begin{document}

\runninghead{The Parthenon Collaboration}

\title{Parthenon -- a performance portable block-structured adaptive mesh refinement framework}

\author{
  Philipp Grete\affilnum{1} \affilnum{2},
  Joshua C. Dolence\affilnum{3} \affilnum{4},
  Jonah M. Miller\affilnum{3} \affilnum{4},
  Joshua Brown\affilnum{5} \affilnum{6},
  Ben Ryan\affilnum{3} \affilnum{4},
  Andrew Gaspar\affilnum{5},
  Forrest Glines\affilnum{2},
  Sriram Swaminarayan\affilnum{5},
  Jonas Lippuner\affilnum{3} \affilnum{4},
  Clell J.~Solomon\affilnum{7},
  Galen Shipman\affilnum{5},
  Christoph Junghans\affilnum{5},
  Daniel Holladay\affilnum{5},
  James M.~Stone\affilnum{8}, and
  Luke F.~Roberts\affilnum{3}
}

\affiliation{
\affilnum{1}Universit\"at Hamburg, Hamburger Sternwarte, Gojenbergsweg 112, 21029, Hamburg, Germany\\
\affilnum{2}Department of Physics and Astronomy, Michigan State University, East Lansing, MI 48824, USA\\
\affilnum{3}Computational Physics and Methods, Los Alamos National Laboratory, Los Alamos, NM 87545, USA\\
\affilnum{4}Center for Theoretical Astrophysics, Los Alamos National Laboratory, Los Alamos, NM 87545, USA\\
\affilnum{5}Applied Computer Science, Los Alamos National Laboratory, Los Alamos, NM 87545, USA\\
\affilnum{6}National Center for Computational Sciences, Oak Ridge National Laboratory, Oak Ridge, TN, 37830, USA\\
\affilnum{7}Eulerian Codes, Los Alamos National Laboratory, Los Alamos, NM 87545, USA\\
\affilnum{8}School of Natural Sciences, Institute for Advanced Study, Princeton, NJ 08544, USA\\
}

\corrauth{Philipp Grete}

\email{pgrete@hs.uni-hamburg.de}

\begin{abstract}

On the path to exascale the landscape of computer device architectures and
corresponding programming models has become much more diverse.
While various low-level performance portable programming models are available,
support at the application level lacks behind.
To address this issue, we present the performance portable block-structured adaptive mesh
refinement (AMR) framework \parthenon, derived from the well-tested and widely used \athena astrophysical magnetohydrodynamics code, but generalized to serve as the foundation for a variety of downstream multi-physics codes.
\parthenon adopts the \kokkos programming model, and provides various levels of abstractions from multi-dimensional
variables, to packages defining and separating components, to launching of parallel compute kernels.
\parthenon allocates all data in device memory to reduce data
movement, supports the logical packing of variables and mesh blocks to reduce
kernel launch overhead, and employs one-sided, asynchronous MPI calls to reduce
communication overhead in multi-node simulations.
Using a hydrodynamics miniapp, we demonstrate weak and strong scaling
on various architectures including AMD and NVIDIA GPUs, Intel and AMD x86 CPUs, {IBM Power9 CPUs}, as well as
Fujitsu A64FX CPUs.
At the largest scale {on Frontier (the first TOP500 exascale machine)}, the miniapp reaches a total of {$1.7\times10^{13}$ zone-cycles/s
on 9{,}216 nodes (73{,}728 logical GPUs) at $\approx92\%$} weak scaling parallel efficiency
(starting from a single node).
In combination with being an open, collaborative project, this makes \parthenon
an ideal framework to target exascale simulations in which the downstream developers
can focus on their specific application rather than on the complexity of
handling massively-parallel, device-accelerated AMR.

\end{abstract}


\maketitle

\section{Introduction}

Many open problems in physics involve vastly varying length- and time-scales.
Some examples, drawn from astrophysics, include 
the deposition and redistribution of energy from active galactic nuclei \citep{meeceTriggeringDeliveryAlgorithms2017,glinesTestsAGNFeedback2020,prasadEnvironmentalDependenceSelfregulating2020,bourneAGNJetFeedback2021}
relativistic accretion flows around compact objects \citep{Ryan_2018,MillerRyanDolence170817,Miller_2020,ressler2020}, 
the in-spiral and merger of neutron stars and black holes \cite{alcubierre2008introduction, Miller_2016},
and, more generally, turbulence simulations \citep{Federrath2021,Grete2021tension}.

From a computational point of view, simulating these problems involves solving (various
types of) partial differential equation -- often on a structured grid using finite volume
or finite difference methods.
However, given the physical scale separation these problems typically cannot be globally
represented in simulations -- even on the next generation, exascale supercomputers.
One option to make these kind of simulations feasible is the use of (adaptive) mesh
refinement (AMR), i.e., a mesh that increases the spatial resolution in regions of interest.
{
  AMR frameworks using varying refinement approaches have successfully been used for
  many years.
  These include refinement based on individual cells, e.g.,
  in \ramses \citep{Ramses} or \rage \citep{Gittings_2008},
  based on separate patches (of arbitrary shape and size),
  e.g., by \citet{BergerColella} implemented in \enzo \citep{Enzo2019} and \pluto \citep{Mignone2011},
  or based on blocks of fixed size, e.g., as in \paramesh \citep{Paramesh}.
  With respect to parallelization all these ``legacy'' frameworks are primarily concerned with handling the mesh
  (and its refinement) across multiple nodes in parallel, see, e.g., \citep{DUBEY20143217} for a
  comparative review.
  Given that they were developed prior to the broad availability of accelerators/GPUs,
  the additional levels of parallelism and memory hierarchy provided by these devices
  are typically not leveraged.
  This prevents an efficient use of those frameworks on many 
  next generation, exascale supercomputers.
}

{From a technical point of view,} achieving sustained application-level exascale performance will require maximizing concurrency
throughout the application while simultaneously minimizing the impact of data movement within the
system. Both issues will be significantly more challenging at exascale than they are on today’s petascale
systems: Amdahl’s law will require ever more levels of parallelism to be exploited in applications to
remove or hide even small sequential bottlenecks.
At the same time technological trends will continue to increase the
expense of data movement relative to compute for most applications
{
  as well as introduce more dynamic performance characteristics due to power capping and highly tapered network topologies.
  An additional challenge is that applications will need to achieve this level 
of performance on two or more radically different system architectures, as typified by the current Summit (IBM/Nvidia) and Frontier (AMD), and future  El Capitan (AMD) and Aurora (Intel) systems.}
These requirements are pushing
applications to consider new programming approaches such as {additional hardware abstraction layers, and/or} compositions of task-based and data parallelism. 

{In general, the combination of accelerated nodes (with large amounts of device memory
  and different architectures) and the complexity of AMR introduces new compuational
  challenges.
 For example, handling many (even up to thousand of) blocks per device with even more
 compute kernels -- especially when small block sizes are involved -- can result
 in significant overheads both with respect to managing the mesh hierarchy as well
 as with respect to the cumulated kernel launch latency.
 }

{To address these challenges,}
we introduce the performance portable block-structured adaptive mesh refinement
framework \parthenon.
It is built on the basis of \athena \citep{athenapp} and \kathena \citep{kathena},
and hides the complexity of AMR {and device computing}
in downstream codes by providing high-level abstractions.
These high-level abstractions not only pertain to the handling of the mesh
{and its data} but also address computational complexity,
{such as parallel execution.}
To exploit on-node data parallelism, \parthenon internally uses the performance portability
programming model \kokkos~\citep{Kokkos2014,Kokkos2021}.
This way \parthenon inherits the \kokkos capability to target various device architectures
{using a single source code and programming model.
To further increase data parallelism, \parthenon also supports various levels
of logical packing of data structures such as variables or even entire blocks,
which are always allocated in device memory to minimize data transfer.}
{To exploit inter-node parallelism, \parthenon internally uses asynchronous, one-sided
(GPU-aware) MPI calls using buffers located in device memory.}

Naturally, the \parthenon collaboration is not the only collaboration who has identified
the various numerical and computational issues of {``next generation'' AMR frameworks.}
{For example, \amrex \citep{Zhang2021} shares many design decisions with \parthenon
  including data containers and abstraction for parallel regions.
  Key differences to \parthenon are the more flexible mesh structure in \amrex (at the cost
  of increased complexity) and a self-contained performance portability layer rather
  than relying on an external library such as \kokkos.
  Another example is \uintah \citep{Holmen2017}, which, as a legacy 
  asynchronous many-task runtime system for block-structured AMR,
  also adopted \kokkos internally
  as performance  portability layer below an intermediate abstraction layer.
  While \parthenon also offers a flexible, asynchronous tasking system, it is
  operating at the block level whereas \uintah tasks can be more fine-grained
  following a directed acyclic graph.
  However, to our knowledge the impact performance of the interplay of fine-grained
  tasks with (many) kernel launches and large number of blocks per device is still
  an open question.
  This simiarly applies to other asynchronous many-task runtime systems such as
  \charm who also start to incorporate GPU support \citep{CharmGPU}. One framework using AMR built on top of \charm is \quinoa \citep{quinoa} that just started to use GPUs.
  Finally, \gamer is astrophysical, multi-physics code with support for GPU-accelerated 
  AMR \citep{Zhang2018}.
  It differs from \parthenon by being a fully integrated code (physics and mesh) 
  rather than an AMR framework and supporting only CUDA  (i.e., Nvidia GPUs).
  Moreover, in \gamer all data structures are allocated in host memory so that data
  required in compute kernels is constantly transferred back and forth between
  host and device memory.
}

In the following, we first provide a brief background on block-structured AMR and \kokkos
in Sec.~\ref{sec:background} before introducing the key design aspects and features
of \parthenon in Sec.~\ref{sec:design}.
In Sec.~\ref{sec:downstream} we provide an overview of various downstream application that
are built on top of \parthenon including the \phydro miniapp.
The latter is used in Sec.~\ref{sec:results} to present different performance
characteristics of \parthenon pertaining to the packing of variables and blocks as well
as to weak and strong scaling.
In Sec.~\ref{sec:sweng} we describe the software engineering approach taken by the
collaboration.
Finally, we discuss current limitations and future enhancements in Sec.~\ref{sec:limitations}
before we conclude in Sec.~\ref{sec:conclusions}.

\section{Background}
\label{sec:background}

\subsection{Block-structured AMR}

Only a brief summary of the block-structured AMR algorithm adopted by \parthenon is given in what follows, a complete description is given in \cite{athenapp}.  Individual cells that span the computational domain are grouped into a regular array of subvolumes termed MeshBlocks.  Data associated with the cells on a given \texttt{MeshBlock} are stored as N-dimensional arrays. \parthenon provides infrastructure for AMR with both cell- and face-centered data.  The size of these arrays must be the same on all MeshBlocks, and moreover the overall domain must contain an integer number of \texttt{MeshBlock}s in each dimension. However, the number and size of individual \texttt{MeshBlock}s tiling the computational domain is arbitrary.

The \texttt{MeshBlock}s themselves are arranged into a binary-tree (in 1D), a quad-tree (in 2D), or an oct-tree (in 3D).  Use of a tree greatly simplifies finding neighbors (necessary for communicating boundary conditions), and allows distribution of \texttt{MeshBlock}s across multiple processers using Z-ordering, which helps improve load balancing.

\begin{figure}
\centering
\includegraphics[width=\columnwidth]{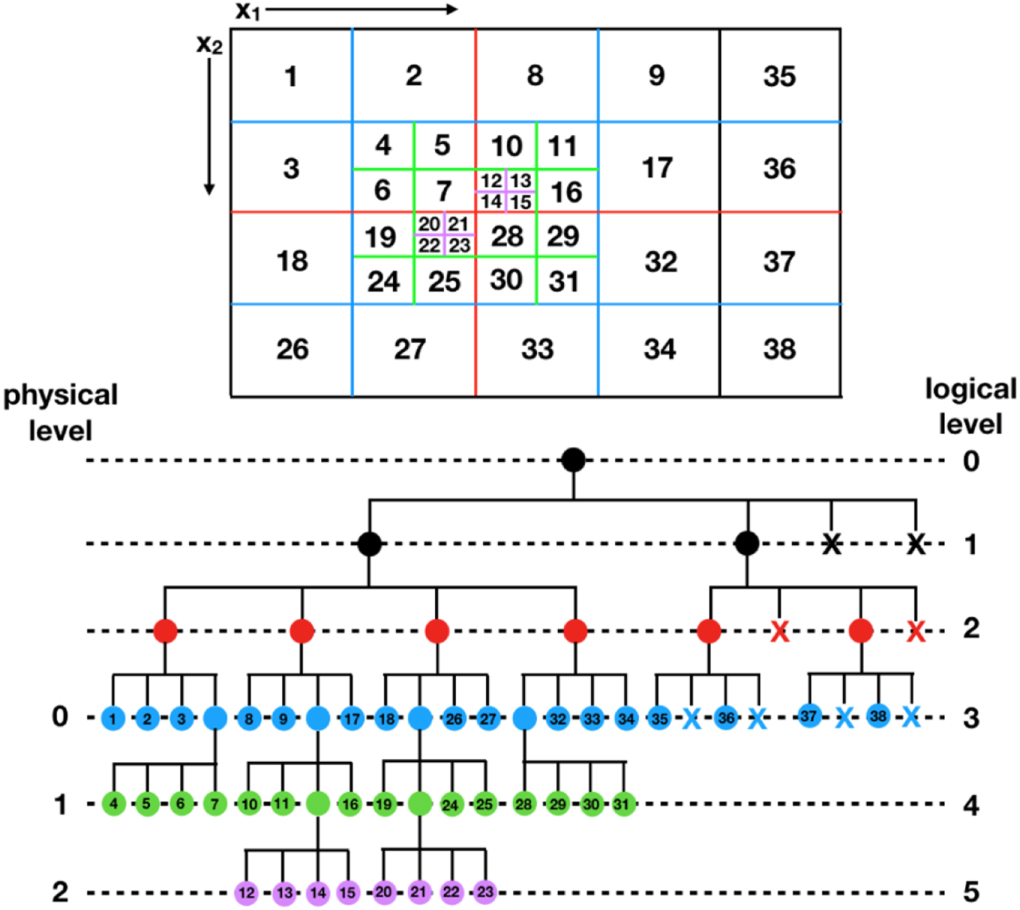}
\caption{
  {
  Labeling of \texttt{MeshBlock}s (top) and their organization into a quadtree (bottom) for an
  example simulation with mesh refinement in two dimensions. 
  Reproduced by permission of the AAS from \cite{athenapp}.
}
}
\label{fig:amrtree}
\end{figure}
For AMR calculations, any number of \texttt{MeshBlock}s can be subdivided into $2^N$ finer \texttt{MeshBlock}s (prolongation), or contiguous blocks of $2^N$ \texttt{MeshBlock}s can be joined into one coarser \texttt{MeshBlock} (restriction), as needed.  Fig.~\ref{fig:amrtree} diagrams how \texttt{MeshBlock}s on a refined grid are stored in the tree.  The tree structure ensures that the neighbors of a \texttt{MeshBlock} can easily be found, even if they are at different levels of the grid hierarchy.  One great advantage of this tree structure-based AMR is that any given spatial location in the domain is covered by one, and only one, \texttt{MeshBlock}.
As a result, only neighbor relationships exist but no spatial parent-child ones.
Thus, except when new \texttt{MeshBlock}s are created or destroyed, prolongation and restriction is required only when data is communicated at \texttt{MeshBlock} boundaries.
However, this approach requires that the entire tree is rebuilt every time (de)refinement is triggered and \texttt{MeshBlock}s are being destroyed/constructed in place.

\subsection{Kokkos}
\kokkos is an open source, performance portable programming model for manycore devices
implemented as a C++ template based library~\citep{Kokkos2014,Kokkos2021}.
As such it provides abstractions to leverage hardware features, e.g.,
threading or multi-level memory hierarchies, through various backends.
This allows device-specific optimization at compile time for devices from
various vendors, e.g., using the CUDA backend for NVIDIA GPUs, the HIP backend for
AMD GPUs, or the OpenMP backend for multi-threading on CPUs.

Some of the fundamental abstractions provided by \kokkos include:
\begin{itemize}
\item \textit{Execution Spaces} define where
  (on which device/through which backend) a computational kernel (in practice a
  function object) is executed.
  \item \textit{Execution Patterns} define how individual work items within a kernel are related.
  Examples include \verb=Kokkos::parallel_for= for independent work items that can be handled
  independently in parallel or \verb=Kokkos::parallel_reduce= to execute a parallel reduction
  over all work items.
  \item \textit{Execution Policies} allow control over how a parallel region is executed.
  They can be simple, such as a \verb=RangePolicy= that correspond to a single
  one-dimensional index for each work item, as well as nested loops, or they can be complex
  descriptions through hierarchical parallelism to control the grouping of threads and
  individual threads.
  \item \textit{Memory Spaces} define where data is stored, e.g., on the host or in device memory, or
  even in cache-type memory (where supported by hardware).
  \item \textit{Memory Layout} allow to specify how data is stored, i.e., how multi dimensional
  indices are mapped to memory locations.
  \item \textit{View}s are the primary data structure provided by \kokkos.
  They correspond to multi dimensional arrays and are parameterized, e.g., by a
  \textit{Memory Spaces} and a \textit{Memory Layout}.
\end{itemize}

\section{Design}
\label{sec:design}
\subsection{Primary design goals}
Many algorithms employed in targeted application domains have
comparatively low arithmetic intensity, e.g., $\mathcal{O}(1)$
floating point operations per byte of data moved for
stencil based calculation.
At the same time, the peak compute power of devices has been increasing
faster than the peak memory bandwidth in recent years and is even
worse for the bandwidth between host memory and device (e.g., GPU) memory.
This results in an ever increasing bottleneck when lots of data needs to be moved.
To circumvent this, \parthenon follows a \textit{device first} or \textit{device resident}
approach in which all work data is allocated in device memory only.
In other words, data movement between host and devices is reduced to a minimum as the
work data used in (expensive) computational kernels is already close to the execution space.

Another goal is to hide complexity from a downstream application point of view.
Similar to \kokkos, which abstracts the complexity of on-node parallel programming, \parthenon
generally provides additional abstractions to hide the complexity of multi node, parallel,
block-structured adaptive mesh refinement.
This includes simplified loop abstractions (i.e., setting many default values in the \kokkos layer)
as well as higher level abstractions such as control over the packing of individual blocks,
communication between nodes via MPI, a tasking infrastructure, or IO, as detailed in the
following sections.
At the simplest level, a downstream application only needs to provide compute kernels
in plain C++ (i.e., no vendor specific backend) that are concerned with data of
a single block and everything else is handled by \parthenon.

Importantly, the underlying access patterns provided by these abstractions need to change
depending on hardware, and must often be tuned for a given problem. To accomodate this
constraint, we expose in our abstraction layers tuning parameters, allowing us to tune
to individual hardware configurations.

Finally, \parthenon is designed with extensibility in mind offering many ``plug-and-play''
interfaces.
This allows for a straightforward addition of many capabilities in downstream codes without
requiring changes in \parthenon itself.
At the same time, this also allows different downstream applications to easily share code
as all downstream features are implemented using those interfaces by construction.

\subsection{Intermediate abstraction layer}

A given set of hardware may require different loop patterns and nested parallelism for optimal performance. For example, an Intel machine parallelized only with MPI may be most performant with a standard C++ for loop, enabled with vectorization pramgas. However, this will obviously not be the case on a GPU. Following the work in \cite{kathena}, we introduce a set of loop abstractions, which we call {\tt parthenon::par\_for} and {\tt parthenon::par\_reduce}. At their simplest, these are thin wrappers around \kokkos parallel dispatch. However, they have a unified interface suited to parthenon loops over meshblocks, regardless of the parallelism pattern used ``under the hood.'' This enables us to swap out \kokkos loops for basic for loops, and calls to the C++ standard library. An example two-dimensional using the basic abstraction might look like
\begin{lstlisting}[language=C++,label=lst:par-for-full,caption={
    Example of a two dimensional \lstinline!for! loop with the \texttt{j} index
    going from \texttt{0} to \texttt{je} and \texttt{i} index from \texttt{0} to \texttt{ie}
    using the basic abstraction provided by \parthenon.
The \lstinline!loop_pattern_tag! controls the \kokkos execution policy
and \lstinline!exec_space! the \kokkos execution space.
}]
parthenon::par_for(
  parthenon::loop_pattern_tag,
  "kernel name", exec_space, 0, je, 0, ie,
  KOKKOS_LAMBDA(const int j, const int i) {
    u(j, i) = ...
  });
\end{lstlisting} 
For ease of use, \parthenon sets several default options, such as the parallel pattern, at
compile time depending on the target architecture.
These are used when the \texttt{par\_for} associated with a \texttt{MeshBlock} are used
as illustrated in the following listing.
\begin{lstlisting}[language=C++,label=lst:par-for-pmb,caption={
    Same as in Listing~\ref{lst:par-for-full} but using the higher level abstraction
    associated with a \texttt{MeshBlock}.
}]
block->par_for("kernel name", 0, je, 0, ie,
  KOKKOS_LAMBDA(const int j, const int i) {
    u(j, i) = ...
  });
\end{lstlisting} 
Note, in constrast to Listing~\ref{lst:par-for-full} neither a \lstinline!loop_pattern_tag!
nor an  \lstinline!exec_space! is set explicitly.

We also introduce an arbitrary rank array abstraction, built on {\tt Kokkos::View}, which we call {\tt ParArrayND}. To support \kokkos layout machinery, we use a six-dimensional {\tt Kokkos::View} as the underlying data structure, and provide a suite of methods for accessing the elements of the array, casting it into a {\tt Kokkos::View}, and getting lower-dimensional slices. This allows us to treat scalar, vector, and tensor variables all in the same way. For example, a three-dimensional array can be allocated as shown in Listing~\ref{lst:pararraynd}.
\begin{lstlisting}[language=C++,label=lst:pararraynd,caption={
    Initializing a three-dimensional \texttt{ParArrayND}.
}]
ParArrayND<double> arr_3d("Array name", 
  n3, n2, n1);
\end{lstlisting} 
The shape is set by \texttt{n3}, through \texttt{n1}. 
Our convention is that the slowest-moving index 
is first in the constructor arguments and higher rank.
However, this depends on the underlying \kokkos memory layout setting.
(We currently assume \texttt{LayoutLeft}.) Our \texttt{ParArrayND} abstraction
supports access operators, where missing indexes are assumed zero, 
slice operators, and access to the underlying {\tt Kokkos::View}, as shown in Listing ~\ref{lst:ParArrayOps}.
\begin{lstlisting}[language=C++,label=lst:ParArrayOps,caption={
    Various operations with \texttt{ParArrayND}.
}]
// Parentheses operator for 
// accessing and setting elements
arr_3d(k,j,i) = value;
// Missing indices are assumed zero
assert(arr_3d(j,k) == arr_3d(0,j,i));
// For lower-rank arrays, 
// extra indices are ignored
assert(arr_3d(l,k,j,i) == arr_3d(k,j,i));
// Returns an array with the second 
// dimension bounded by lower and upper
auto sliced = 
 arr_3d.SliceD<2>(lower, upper);
// Returns a three-dimensional 
// Kokkos View
auto view_3d = arr_3d.Get<3>();
\end{lstlisting} 

Both host and device {\tt ParArrayND} objects are supported, but they default to living in device memory.

\subsection{Packages}

\parthenon is designed to couple multiple disparate components together. To capture this, we introduce \textit{packages}. Each package is an independent functionality built on top of \parthenon, with its own registered variables, physics routines, and tasks. Importantly, packages can \textit{share} variables. In other words, package ``A'' may register a variable and package ``B'' may use it. \parthenon supports dependency tracking between variables registered by packages. A package may register a variable as
\begin{itemize}
    \item Private
    \item Provides
    \item Requires
    \item Overridable
\end{itemize}
A \textit{Private} variable is private to a given package, and lives in the package's namespace. Other packages should not access it. A \textit{Provides} variable is provided by a package, with the intent that other packages may use it. However, the providing package is expected ``own'' the variable. If two packages try to provide the same variable, an error is raised. If a package registers a \textit{Requires} variable, it is stating that it needs this variable to exist, but does not create or manage it itself. If no package provides a required variable, an error is raised. If a package registers an \textit{Overridable} variable, it is stating that it can provide this variable, but will defer to another package, if it provides it.

\begin{lstlisting}[language=C++,label=lst:init1,caption={
    An example package initialization function.
}]
namespace my_package {
auto Initialize(ParameterInput *pin) {
  using SD = StateDescriptor;
  // this pkg object is where we register
  // things like variables
  auto pkg = 
    std::make_shared<SD>("my package");
  // Metadata objects contain 
  // information about variables.
  // This variable is cell-centered, 
  // and provided by this
  // package.
  auto m = 
    Metadata({Metadata::Cell, 
      Metadata::Provides});
  pkg->AddField("My Variable", m);
  // This variable is expected 
  // to exist but not provided.
  m = Metadata({Metadata::Requires});
  pkg->AddField("I need this", m);
  
  return pkg;
}
} // namespace my_package
\end{lstlisting} 

Packages register their variables, as well as global constants within their namespace (called \textit{params}) in a function we call \textit{Initialize}. An example {\tt Initialize} function is shown in Listing~\ref{lst:init1}. All initializations are registered by the parthenon manager object at startup. To tell the code what packages to load, a {\tt ProcessPackages} function must be provided. An example function is shown in Listing~\ref{lst:process}. 

\begin{lstlisting}[language=C++,label=lst:process,caption={
    An example function for adding packages.
}]
using PI = ParameterInput;
using Pin_t = std::unique_ptr<PI>;
Packages_t ProcessPackages(Pin_t &pin) {
  Packages_t packages;
  auto pkg1 = 
    my_package::Initialize(pin.get());
  auto pkg2 = 
    my_other_package::Initialize(pin.get());
  packages.Add(pkg1);
  packages.Add(pkg2);

  return packages;
}
\end{lstlisting} 

Note that although packages create their own variables and provide tasks, these tasks are not automatically called. The tasks must be woven together ``by hand'' by an expert in the driver code. This will be explained in Section \ref{sec:tasks}.

\subsection{Variables}

Variables in \parthenon consist of metadata and data. The data is stored on a per-block basis in a multidimensional {\tt Kokkos::View}. It can live
at cell centers, faces, edges, corners, or not be associated with a mesh entity at all. Although, in the initial \parthenon release, only 
cell-centered and non-mesh-tied variables are fully implemented. Support for the other types of variables will be added in a later release.

All variables in \parthenon must be named. The name is used in simulation output, error messages, and to obtain a handle to the variable data 
from containers (see Sec.~\ref{sec:packing}). This greatly enhances the readability and self-documentation of the code. The name of a variable is 
stored in its metadata along with other important information. The metadata also contains the shape of the variable, i.e.\ if it's a scalar,
vector, or tensor, along with the number of components in each dimension in the case of vectors and tensors. Finally, the metadata contains
a collection of flags that indicate, for example, if the variable is independent or derived, whether it's private, provided, required, 
or overridable (see previous section), if it's advected, if it needs ghost cells filled, if it has fluxes, etc.

The metadata information allows the \parthenon infrastructure to perform certain tasks on variables without needing to understand their
physical meaning. For example, \parthenon can write a restart file that includes only the independent variables, since they are all flagged as such.
When using reflective boundary conditions, \parthenon can reflect the X-component of vector variables in the X-direction, Y-components in the Y-direction,
and so on. Furthermore, the metadata flags are also useful for user provided physics packages. For example, the hydro package can advect all
variables from all packages flagged as advected, without needing to know what those variables are. By setting the {\tt FillGhost} and {\tt WithFluxes} 
metadata flags, the user can control which variables will have their ghost cells filled by \parthenon and which variables will have fluxes buffers
allocated.

Typically, variables are allocated on every block in the entire domain. But for some applications, there may be variables that are only relevant 
in parts of the domain, thus creating opportunities to save both memory and computing resources. For such cases, \parthenon provides sparse 
variables. Sparse variables behave just like ordinary (or dense) variables, with two exceptions: i) Instead of just a name, sparse variables 
have a base name and a sparse ID, and ii) sparse variables are only allocated on some blocks.

Sparse variables are added through \emph{pools}. A sparse pool consists of a base name, a set of sparse IDs, and shared metadata. For each 
sparse ID in the pool (e.g. 1, 4, 10, 11), a sparse variable is created whose name is ``basename\_X'', where ``basename'' is the pool's basename and ``X'' is the
sparse ID. The sparse variables have the same metadata as the pool's shared metadata, except for the shape and Vector/Tensor flags,
which can be set individually per sparse ID. Furthermore, the sparse variables are not allocated on any blocks until the user manually allocates
them on specific blocks or they are advected into a block where they were not previously allocated. They can also be deallocated by the \parthenon
infrastructure if they completely leave a block. The main use case for sparse variables are multi-material simulations where a particular sparse ID
corresponds to a particular material. Currently, only cell-centered variables are supported as sparse variables.

\subsection{Particles}

In addition to the structured multi-dimensional variables (either tied to mesh entities or not) described above, \parthenon also supports particle data structures, called \texttt{Swarm}s. Like variables, swarms combine metadata and data, and are stored on a per-block basis. Swarms hold particle data in a Struct of Arrays pattern; as such, particles that will be iterated over together by the same physics should belong to the same swarm. 

Swarms support a subset of Metadata flags used by variables; Provides or Requires are used by individual packages to share particle data, and None is generically set because particles are not grid-based quantities. A swarm is composed of a set of \texttt{ParticleVariable}s, which store data in 1D \texttt{ParArrayND}s. Each particle variable contains its own metadata; in particular, this metadata is used to specify the datatype of the particle variable, either real or integer. Swarms are always created with \texttt{x}, \texttt{y}, and \texttt{z} real-valued particle variables; additional variables are enrolled by the package creating the swarm. This approach of user-specified data with memory locality provided by the library has been successfully applied in other particle frameworks \citep{Zhang2021, cabana}. 

In general, the particle population will grow and shrink in size over time, particularly on the scale of a meshblock. This can occur both through physics algorithms that create or destroy particles and communication of particles across meshblocks. Swarms manage their memory dynamically; users request the creation of a certain number of particles. Existing empty elements in the particle list are filled in first, and then if necessary the swarm will internally resize its ParticleVariables to accommodate the remaining particles. This resizing procedure proceeds exponentially to limit the number of memory reallocations required; the size of the memory pool grows by factors of 2. Swarms include a Defrag method that deep copies individual particles’ entries to ensure contiguous memory in each particle variable on demand.

Particle communication is handled by non-blocking send and receive calls as in grid-based data communication. During package functions that update particle positions, particles must be checked for whether they have left the meshblock they are currently on. This will be recorded by the swarm, and during the subsequent send and receive calls the off-block particles will be copied to either send buffers for subsequent MPI communication or copied directly onto the receive buffers of blocks on the same MPI rank. The sent particles are deleted from the sending meshblock’s swarm. Receiving meshblocks then copy the particles from the receiving buffers into their own swarm’s particle variables. Only communication to neighboring meshblocks is supported. 

Particle communication between the same meshblocks can be required multiple times per timestep, particularly for algorithms where particles can traverse many meshblocks per timestep. This can be implemented by a separate blocking \texttt{TaskRegion} that is repeatedly called until a global stop criterion is met, as in the provided examples, or through the iterative task list machinery. 

Boundary conditions on particles are applied to all particles marked as being off their meshblock by the internal swarm send and receive tasks. Boundary conditions are implemented through separate polymorphic boundary condition classes for each of the six boundary faces. \parthenon provides periodic and outflow boundary conditions; additional boundary conditions can be implemented by driver applications. 

Particles are not sorted by grid zone below the scale of an individual meshblock. Particle-mesh interactions are handled via \kokkos atomics by the downstream application.

\subsection{Data containers/Packing}
\label{sec:packing}

As discussed above, each package may register its own set of variables. However, it is often useful to loop over all variables, either sparse or dense, with some set of properties such as the need to perform ghost halo exchange. Because launching code on an accelerator comes with some (often significant) latency, it is also often far more performant to bundle work \textit{across} mesh blocks into a single device kernel launch.

To enable this, we implement \texttt{VariablePack}s and \texttt{MeshBlockPack}s.
\texttt{VariablePack}s are objects that collect all desired variables within a single index space. {In the process, indices of higher rank variables (e.g., tensors) are flattened so
that all variables (and their components) can be accessed by a single running index,
typically \texttt{v} in addition to the spatial \texttt{k}, \texttt{j}, and \texttt{i}
indices.
The underlying data structure is a \texttt{View} of \texttt{View}s allowing efficient
access to the existing data on devices.}
Variables {for \texttt{VariablePack}s} can be selected via metadata  tags registered by a given package, or by name.
\texttt{MeshBlockPack}s do the same, but also gather variables from some number of mesh
blocks on a given MPI rank.
{This results in an additional, fifth flattened index, typically notated by \texttt{b}.}
The optimal number of mesh blocks to gather is hardware and problem dependent, and so may be set at runtime, see Sec.~\ref{sec:res-pack-sizes} for some example results. To expose these packing mechanisms, as well as relevant metadata used in a given physics kernel, we implement the \texttt{MeshBlockData} and \texttt{MeshData} data structures. These objects have methods to generate pack objects and also automatically cache the relevant packs from cycle to cycle. The \texttt{MeshBlockData} and \texttt{MeshData} objects also expose accessors for variables, grid shape information, and parameters set by individual packages.
{Overall, this allows efficient access to all data of an arbitrary number of variables on an arbitrary number of blocks through tight, 5-dimensional loops.}

\subsection{Boundary communication}
\label{sec:boundaries}
Two important strategies to achieve a high parallel efficiency across multiple ranks
are implemented in \parthenon.

First (and more general), all communication buffers can be exchanged asynchronously
by using one-sided, asynchronous MPI calls.
Moreover, each \texttt{Variable} uses its own MPI handle so that individual
\texttt{Variables} can also be communicated independently.
This also applies to flux correction for multi level meshes.
A typical driver to solve equations in conservative form 
implements several boundary communication related tasks that are split
on purpose.
These tasks include
a) initializing/resetting the individual MPI handles,
b) starting and receiving flux correction (with mesh refinement enabled
after calculating block local fluxes), 
c) filling communication buffers with the updated data (e.g., after calculating the
flux divergence), 
d) start sending communication buffers (via \verb=MPI_Start=)
e) fill ghost cells from buffers already received.
These tasks can be run for individual blocks and variables and, thus, allow
to hide communication related walltime (e.g., latency) behind computations.
In other words, while buffers of some blocks are filled in a compute kernel
executed on a device, already filled buffers of other blocks can already be
communicated in parallel in the background.

\begin{figure}[tb]
\centering
\includegraphics[width=\columnwidth]{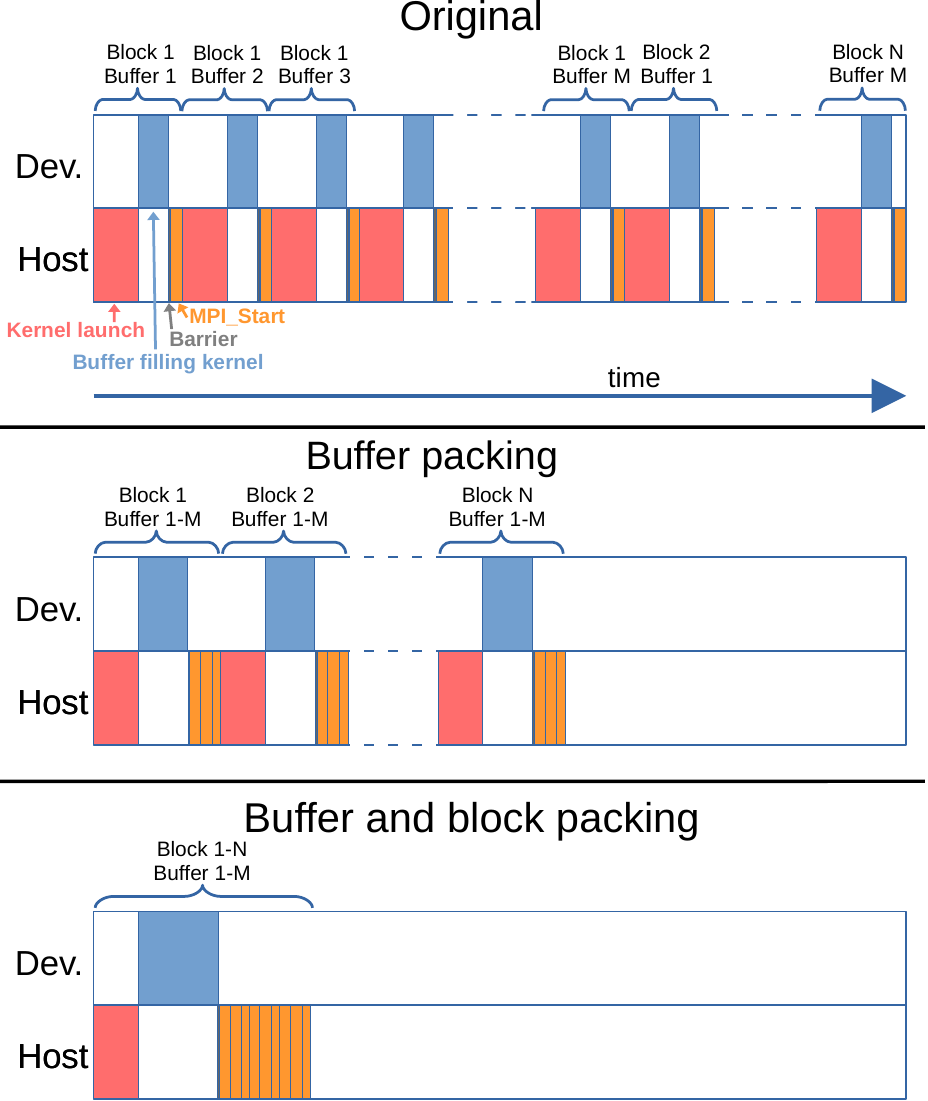}
\caption{{Illustration of the buffer and block packing machinery in \parthenon.
  (top) In the original refactoring from \athena each communication buffer of each
  block is packed separately and sequentially with the runtime of the kernel typically
  being smaller than the kernel launch overhead itself.
  (middle) With buffer packing all communication buffers of a single block are
  filled in a single kernel (with slightly larger runtime -- but more parallelism inside
  the kernel).
  (bottom) With buffer packing and block packing all buffers of all blocks in pack
  (number of blocks per pack is a runtime parameter) are filled in a single kernel
(allowing for even more parallelism).}
} 
\label{fig:buf-block-pack}
\end{figure}
The second (and more specific to GPU-accelerated simulations) strategy is
filling more than one communication buffer in a single kernel.
In \athena each buffer is filled independently in small pack and unpacking routines.
However, the work done in these buffer filling kernels is very small, e.g., just copying
8 numbers for a corner buffer of a 3D block with 2 ghost zones in each direction,
making the actual kernel runtime
significantly smaller than the kernel launch time (typically a couple of $\mu s$).
Given that some vendor APIs (e.g., when running with the CUDA backend) are inherently
serial for launching kernels, no significant performance increase can be expected even
when multiple kernels can be executed in parallel on the device.
For this reason, we implemented a flexible ``fill-in-one'' approach that allows
us to fill all buffers of one or more \texttt{Variables} on one or more blocks in a
single kernel{, see Fig.~\ref{fig:buf-block-pack} for an illustration.}
The performance in practice of this approach is shown in Sec.~\ref{res:pack-in-one}. 

With the block structured AMR adopted in \parthenon, prolongation and restriction of data
only occurs during communication of data between neighboring \texttt{MeshBlock}s at different levels of
refinement, and therefore these steps are functionally part of the boundary communication design.
Data sent from fine-to-coarse levels are first restricted and then communicated to reduce
message sizes.  Data sent from coarse-to-fine are packed into special coarse buffers on the
target \texttt{MeshBlock}.  Once all communication has completed, the data in these coarse buffers are then
interpolated (prolongated) to the fine resolution.  Details of {the multi-level
communication and interpolation}
algorithms for cell- and face-centered data are given
{in Sec.~2.1.3 and 2.1.5 of \citet{athenapp}.}
Again, in order to reduce the number kernel launches restriction is now handled within the
``fill-in-one'' machinery in contrast to \athena where each restriction is a separate kernel.

{Finally, contrary to the \athena design each \texttt{Variable} uses a unique \mpi 
communicator rather than the default communicator and individual buffers use \mpi tags
created sequentially rather than globally.
The key advantage is to circumvent the minimum upper bound of at least 32{,}767 defined
by the \mpi standard.
This bound is easily reached when running 3D mesh refinement simulations with small block
sizes on modern devices where a single rank can (computationally) easily handle 100s to
1000s of blocks.
}

\subsection{Load balancing and mesh refinement}
When new \texttt{MeshBlock}s are created or destroyed as part of the AMR, load balancing of the
resulting workload across devices becomes important.  Following the strategy in \athena (see section 2.1.6 in \cite{athenapp}),
in \parthenon \texttt{MeshBlock}s are redistributed across nodes whenever mesh refinement occurs and the tree
is rebuilt.  Generally some fractions of the \texttt{MeshBlock}s on each device will have to be moved to neighbors to
achieve good balance. Nevertheless, the increase in performance associated with good load
balancing outweighs the overhead of this communication.  Note that mesh derefinement is only allowed periodically {(controlled by a runtime parameter)} 
to prevent regions very close to the criterion from refining and then derefining on subsequent cycles.

{For performance, the new tree structure is always rebuilt first and 
  that information is used to determine the meshblock distribution across ranks.
Thus, only afterwards the tree is populated with data either by a) moving pointers to
\texttt{MeshBlock} objects for same-level, same-rank blocks from the old to the new tree,
b) by creating or destroying blocks for same-rank, (de)refined blocks,
or c) by sending meshblock data to a different rank.
For the latter, the data transfer is optimized for size, i.e., if blocks can be derefined
on the sending rank, this is done first before sending the data, and, similarly, if the
block needs to be refined, the original data is being sent and the refinement occurs on
the receiving rank.}

\subsection{IO}

\parthenon uses (parallel) \hdf to read and write simulation data.
An arbitrary number of different outputs can be defined for a given simulation that can
differ in the time interval for writing output, the variables contained, the
precision (single or double precision floating point numbers) and the compression level.
The latter is also enabled through the \hdf library and allows for inline compression, which
is particularly useful for sparse variables.
Several environment variables are processed by \parthenon for a fine grained control
of both \hdf parameters as well as MPI-IO parameters.
{For performance, data locality, and (optional) compression \hdf chunking is used
where each chunk corresponds to the meshblock data of a \texttt{Variable} component.}
The special ``restart'' output type forcibly includes all variables with the \texttt{Independent}
or \texttt{Restart} \texttt{Metadata} flags and write output in double precision.
They allow for a simulation to be restarted in a bitwise identical manner.
Moreover, when restarting a simulation a different number of \mpi ranks can be chosen, e.g.,
to adapt to a changing number of \texttt{MeshBlock}s when using AMR.
{This is naturally handled by the load balancing mechanism as the tree is being rebuilt
upon restarting a simulation.}

\parthenon also automatically writes \texttt{xdmf} files along the data files, which allows
external (analysis) tools such as \paraview or \visit to directly read the output data.
Finally, a \yt frontend is currently being reviewed and expected to be merged soon.

\subsection{Tasks and reductions}
\label{sec:tasks}
\begin{figure}
\centering
\includegraphics[width=\columnwidth]{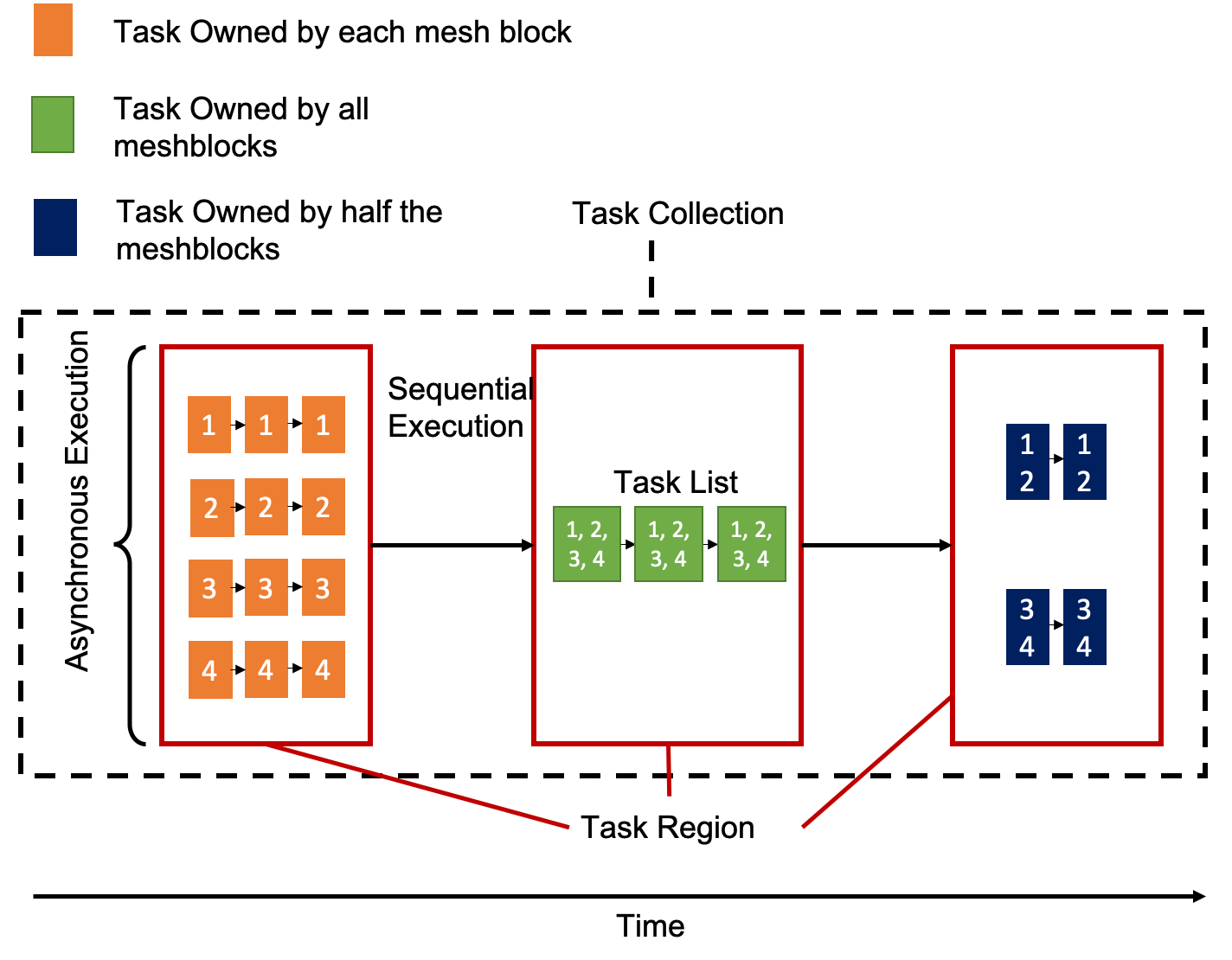}
\caption{Tasks are organized into regions which are in turn organized into collections.
Task regions within a collection are executed sequentially and each task region can have a different
granularity. The illustrated task collection is composed of three task regions, controlling the
execution of tasks on four \texttt{MeshBlock}s, indicated by the numbers.
The first region launched (potentially) concurrent tasks on each \texttt{MeshBlock}, where the
dependencies of a given task can only be other tasks that operate on the same \texttt{MeshBlock}.
Once all tasks in the region are complete, the execution moves to the next region where three
tasks are launched that each operate on all four \texttt{MeshBlocks} simultaneously.
Finally, once these are complete, execution moves to the final region which defines tasks that operate
on subsets of \texttt{MeshBlock}s. In this way, task granularity is controlled at the
task region level and overall execution is controlled at the collection level.
}
\label{fig:tasks}
\end{figure}

\parthenon provides a simple infrastructure for exploiting task-based parallelism.  Tasks are organized hierarchically in \texttt{TaskCollection}, \texttt{TaskRegion}, and \texttt{TaskList} objects.  In typical usage, applications build and execute a \texttt{TaskCollection} object that encapsulates each stage of a calculation, which might correspond to a time step or even a single Runge-Kutta integrator stage.  Each \texttt{TaskCollection} is made up of one or more \texttt{TaskRegion}s, each of which contains one or more \texttt{TaskList}s.  At the lowest level of the hierarchy, tasks are added to \texttt{TaskList} objects by capturing the function to be executed, all of its arguments, and any dependencies that must be executed before the task can be launched.  Tasks in a \texttt{TaskList} all operate on data at the same granularity, be that the data on a single `MeshBlock` or data across multiple `MeshBlock`s.  Tasks in different \texttt{TaskList} objects within a \texttt{TaskRegion} can be executed concurrently, but \texttt{TaskRegion}s are serialized within a \texttt{TaskCollection}.  Fig.~\ref{fig:tasks} illustrates these relationships.

Many algorithms require the ability to do global reductions.
In a task-based environment where each rank may be executing multiple tasks lists
operating on independent sub-domains, orchestrating these reductions is nontrivial.
\parthenon provides task-based global reductions for typical datatypes encountered in downstream
applications such as plain integers or floating point data, \lstinline!std::vector!s thereof, and
\lstinline!Kokkos::View!s or \lstinline!parthenon::ParArrayND!s.
Reductions are realized by updating a shared rank-local variable from individual tasks in each
\texttt{TaskList}.
Those tasks are marked as a shared dependency within that \texttt{TaskRegion}.
Only after all tasks with the shared dependency are completed a
non-blocking MPI reduction operation is called from a single task on each rank.

\subsection{Application driver}
In \parthenon-based applications, a driver orchestrates the execution of a computation by building and executing collections of tasks, calling I/O functions as needed, and calling into the load balancing and AMR capabilities, if desired.  \parthenon provides a basic set of driver classes from which applications can derive.

At the most basic level, the \texttt{Driver} class gives access to the mesh and I/O capabilities, but assumes nothing about the type of calculation being performed.  Downstream applications must define an \texttt{Execute} function that encapsulates the entirety of the control flow and execution.  The \texttt{calculate\_pi} example demonstrates a capability that derives from \texttt{Driver}, namely one that approximates the value of $\pi$ using AMR.

Deriving from \texttt{Driver}, the \texttt{EvolutionDriver} is appropriate for applications that evolve a solution through time.  In this case, \texttt{Execute} is already defined.  When applications derive from \texttt{EvolutionDriver} they must provide a \texttt{Step} function that is responsible for evolving a solution through a timestep.  The \texttt{EvolutionDriver} calls this \texttt{Step} function from within a loop that executes until a specified amount of simulated time has elapsed, calling the I/O, load balancing, and AMR capabilities as appropriate.

Finally, \parthenon provides a \texttt{MultiStageDriver} which derives from the \texttt{EvolutionDriver}, defining the \texttt{Step} function as appropriate for a multi-stage Runge-Kutta integration.  In this case, the downstream application need only define a \texttt{MakeTaskCollection} function which must build the \texttt{TaskCollection} object appropriate for a single stage of the integration.  The \texttt{advection} example demonstrates the usage of this driver class.

\subsection{Machine dependent build configuration}
While the hardware environment becomes more heterogeneous (requiring performance portable
approaches), the software environment similarly adapts and becomes more heterogeneous.
For example, custom launchers like \lstinline=jsrun= on OLCF's Summit are developed and 
used to allow for an appropriate mapping of hardware resources to processes for parallel
execution.
At the same time, the user has to choose a suitable mix of compiler, communication,
and potentially offloading libraries for configuring, compiling and running a code.

For ease of use, \parthenon ships with so-called machine configuration files for
various supercomputers.
These files contain default values, e.g., architecture specific flags or parallel
launch commands, as well as a recommendation for the environment modules to load.
The configurations are regularly tested and updated to reflect the latest
software environment provided on a system.
This allows (new) users to readily compile and run the test suite without being
bothered by machine specific details.

\section{Downstream applications}
\label{sec:downstream}
\subsection{\phydro}

\phydro\footnote{\url{https://github.com/parthenon-hpc-lab/parthenon-hydro}} is
a minimal implementation of algorithms solving the Euler equations.
In contrast to the examples included in the \parthenon repository, which are
mainly used to demonstrate and/or test individual features, \phydro is
considered a fully-fledged miniapp consisting of just $\approx$1400 lines of C++ code total.
Its main purposes are to both illustrate a possible use of various
\parthenon features combined in practice as well as an external integration and
performance test.
\phydro supports 1D, 2D, and 3D compressible hydrodynamics on uniform and
(static and adaptive) multi level meshes.
Given \parthenon's \athena origins, \phydro is also based on a subset of the
algorithms implemented in \athena.
More specifically, \phydro uses a second-order method consisting of
a two-stage Runge-Kutta integrator, piecewise linear reconstruction and
HLLE Riemann solver.
For illustration purposes following three problem generators are implemented:
a linear wave (which is also used to illustrate automated convergence testing by
reusing the \parthenon infrastructure externally), a spherical blast wave, and
a Kelvin-Helmholtz instability to illustrate adaptive mesh refinement.
There are no plans to further extend ``physics'' capabilities of \phydro
with the exception of demonstrating new features in \parthenon as
hydrodynamics is also supported by other, more feature rich downstream
applications such as \athenapk.

\subsection{\athenapk}

\athenapk (Athena-Parthenon-Kokkos) is a general purpose astrophysical
magnetohydronamics code which serves as a performance-portable, AMR-capable
conversion of \athena \citep{athenapp}. It implements the hydrodynamics
solvers from \athena and supplemented them with a divergence cleaning
magnetohydrodynamics solver based on
\citet{dednerHyperbolicDivergenceCleaning2002}.

\begin{figure}[tb]
\centering
\includegraphics[width=\columnwidth]{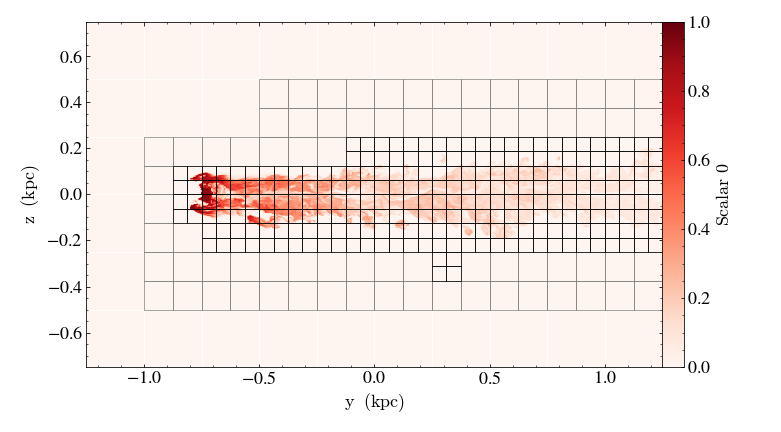}
\caption{\athenapk example: Passive scalar concentration in a supersonic cloud crushing
  simulation with magnetic fields, optically thin radiative 
cooling, and mesh refinement configured to follow the cloud material (as passive scalar).
} 
\label{fig:athenapk}
\end{figure}

At present, \athenapk is used for simulations of magnetized galaxy clusters with
feedback from active galactic nuclei, cf.,
\citet{meeceTriggeringDeliveryAlgorithms2017,glinesTestsAGNFeedback2020,prasadEnvironmentalDependenceSelfregulating2020},
cloud crushing in galatic outflows, and magnetohydrodynamic turbulence.
To support these applications additional features implemented include
various Riemann solvers, passive scalars, tabulated cooling, and (an)isotropic thermal conduction
with support for 2nd-order Runge–Kutta–Legendre based super-time-stepping \citep{Meyer2014},
see Fig.~\ref{fig:athenapk}
for an example multi-physics simulation with AMR. 

Development of \athenapk is public and contributions are welcome\footnote{
  \url{https://github.com/parthenon-hpc-lab/athenapk}}.

\subsection{\phoebus}

\begin{figure}[tb]
\centering
\includegraphics[width=\columnwidth]{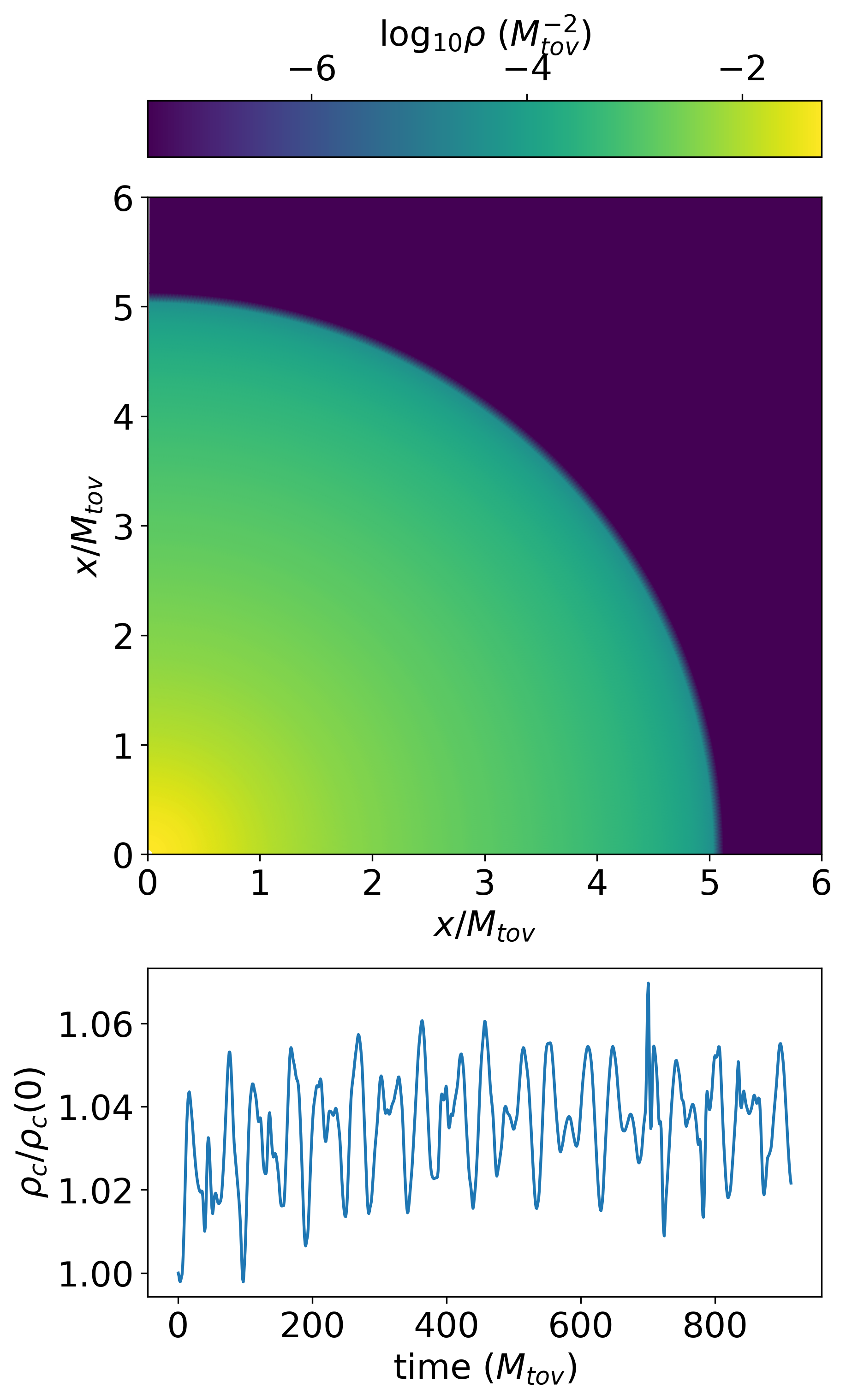}
\caption{Self-gravitating compact star as evolved in \phoebus (top) and oscillations in the central density (bottom). The star is stable for many cycles, and the oscillations match the expected quasinormal mode structure for a non-rotating neutron star.} 
\label{fig:tov}
\end{figure}

\phoebus\footnote{\url{https://github.com/lanl/phoebus}} is a general relativistic neutrino radiation magnetohydrodynamics code, designed for modeling compact binary mergers and their aftermath. It uses the Valencia formulation of general relativistic hydrodynamics \citep{Valencia1,Valencia2}, with constrained transport for magnetic fields. Currently the cell-centered formulation of \cite{Toth} is utilized, but face-centered fields will be leveraged once the underlying data structures are implemented in \parthenon. On the radiation side, \phoebus implements Monte Carlo transport as in \cite{nubhlight}, and a novel hybrid scheme first presented in \cite{MOCMC} is in development. Currently \phoebus implements both arbitrary fixed spacetimes as well as self-gravity under the monopole approximation. Full dynamical numerical relativity is a planned improvement. \phoebus carries with it several challenges: a general relativistic background carries with it a very large state vector, with O(100) variables; the fluid primitives are no longer trivially solve-able from the conserved variables, and must be found via numerical root finding; the method requires the interweaving of grid and particle variables; and the general relativistic equations themselves are complicated and compute intensive. 
A code paper for \phoebus is in preparation.

Figure \ref{fig:tov} shows an example problem run in \phoebus: a non-rotating neutron star. The top panel shows the density in a poloidal slice. {Any small perturbation excites the natural} oscillation modes in the star, shown in the bottom panel, where we plot the central density, normalized by its value at the initial time. {These modes match those predicted from perturbation theory \citep{YoshidaQuasinormal} and presented in numerical tests in, e.g., \cite{EinsteinToolkit}}.

\begin{figure}[tb]
\centering
\includegraphics[width=\columnwidth]{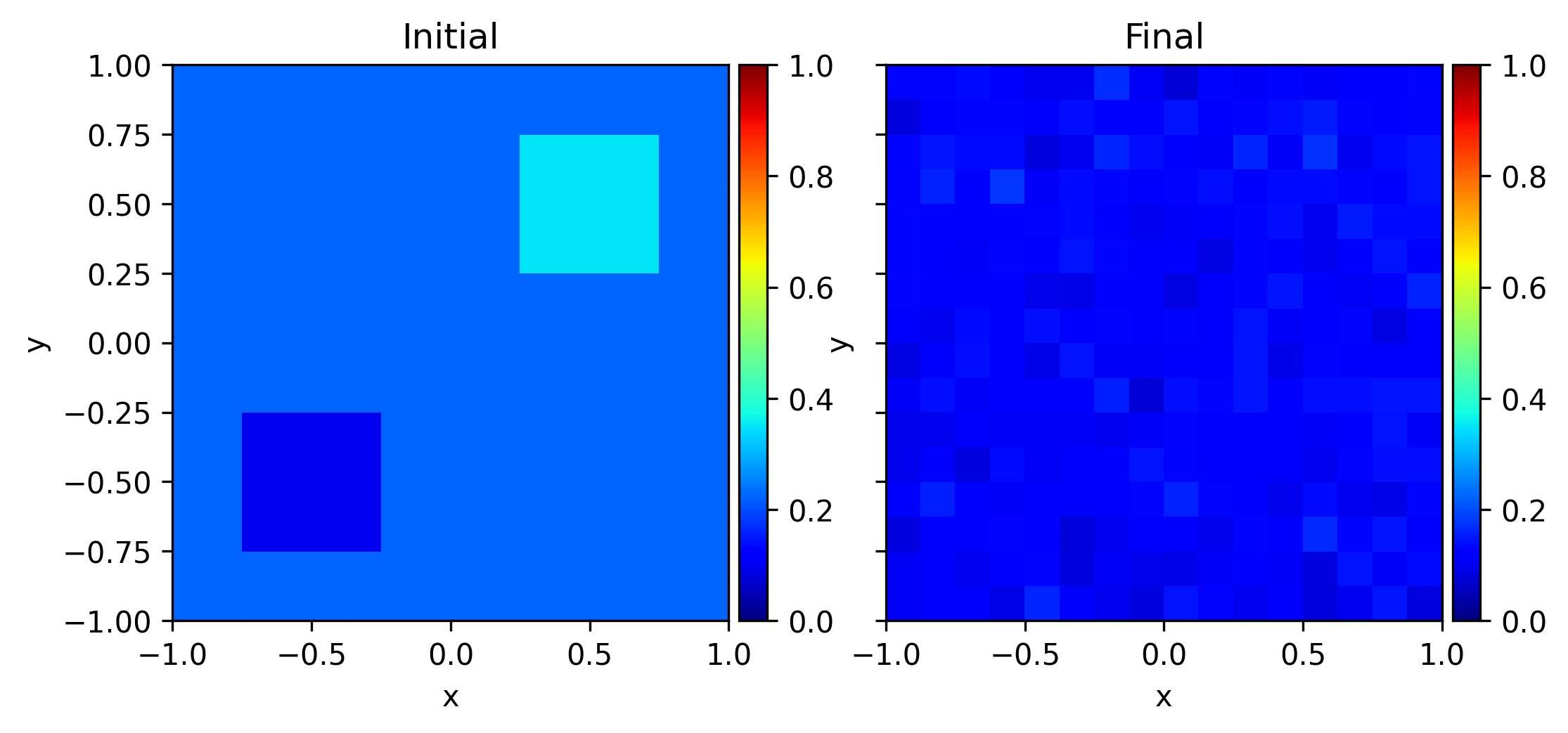}
\caption{Initial and final electron fraction material states of the leptonization neutrino transport problem. Electron fraction-dependent emissivities act to equilibrate the electron fraction across the simulation domain from the inhomogeneous initial conditions. The mean electron fraction of the material is lower at the final time due to the presence of neutrinos.} 
\label{fig:leptoneq}
\end{figure}

Figure \ref{fig:leptoneq} shows another \phoebus example problem using Monte Carlo neutrino transport, leveraging the \parthenon particles infrastructure. In this problem, an initially inhomogeneous electron fraction, the ratio of electrons to baryons, of the background material is homogenized by neutrino emission, transport, and absorption (neutrinos transport lepton number). Inside \phoebus, we use the \singularityeos \citep{MillerSingularityEOS} library for a realistic equation of state \citep{Fornax} and the \singularityopac \citep{MillerSingularityOpac} library for realistic opacities \citep{OConnorOtt2010, Steiner+2013}. Singularity libraries provide production-quality data in a performance-portable way.

\subsection{\riot}

\riot is a LANL-based multiphysics code designed to emulate a subset of the physics in the \rage code \citep{Gittings_2008} to allow for comparisons of cell-based and block-based AMR approaches.  Currently it includes multi-material, compressible hydrodynamics with a pressure-temperature equilibrium mixed-cell closure, gray radiation diffusion, a sub-grid turbulence model, thermonuclear reactions, and high-explosives models.  \riot makes heavy usage of \parthenon's sparse data type to represent material based state variables.

\begin{figure}[tb]
\centering
\includegraphics[width=\columnwidth]{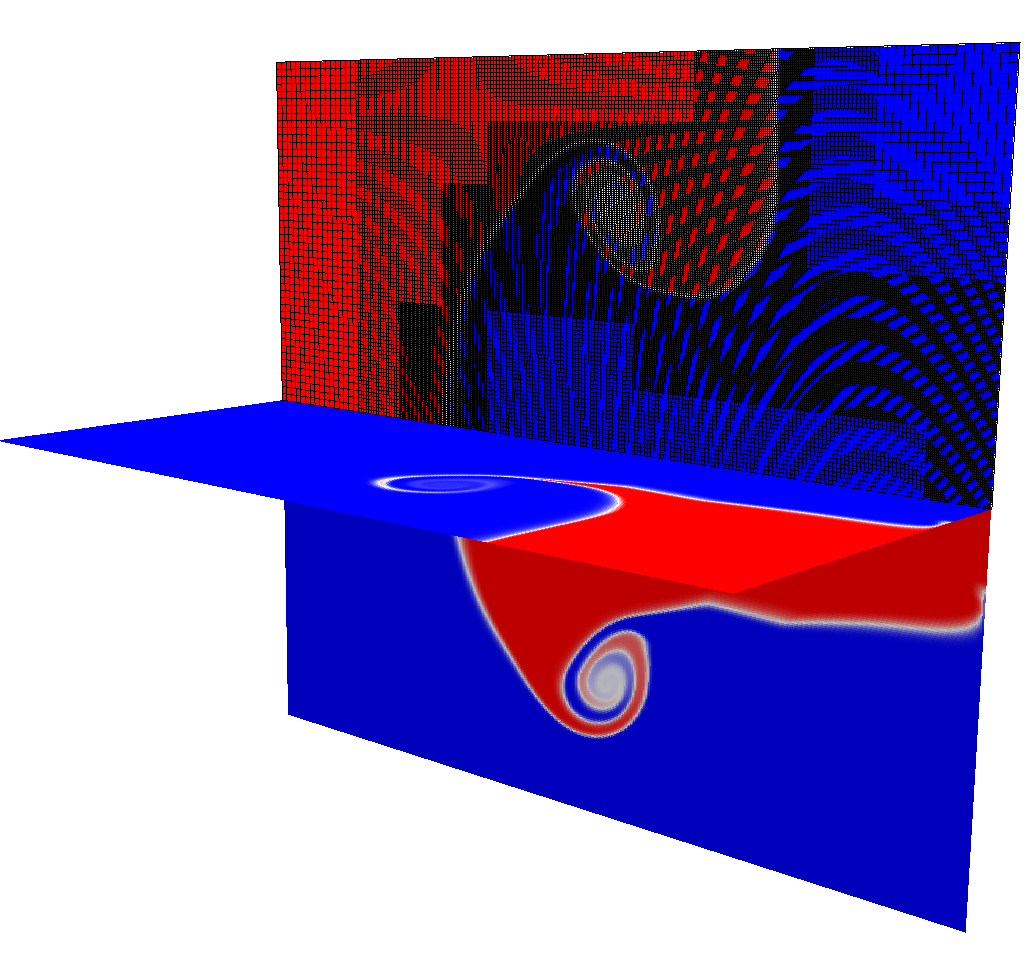}
\caption{Slices of material volume fractions in the 3D three material triple-point problem at t=5.0.  \parthenon's mesh infrastructure enables \riot to maintain high-resolution around material interfaces, as shown in the top slice.} 
\label{fig:triple}
\end{figure}

Figure \ref{fig:triple} shows results from a classic three material test problem called triple-point \citep{triple_point}.  At one end of the domain, an ideal gas at high pressure drives a shock into two distinct ideal gases that differ in their adiabatic index $\gamma$.  The flow develops vorticity that leads to a well-developed roll-up.  The problem was solved in a 3D geometry by revolving the traditional 2D setup about the y-axis and made use of three levels of refinement triggered by material interfaces.  The figure shows slices of volume fraction for each material where blue indicates the absence of the material and red indicates a pure material, with white indicating material mixing.  On the top slice, we also show the AMR grid to indicate how \parthenon adapts the mesh to accommodate the evolving and nontrivial geometry of the materials.

\section{Results}
\label{sec:results}
Unless noted otherwise, all result presented in this section were obtained using 
\phydro (changeset \texttt{52fa13c} with included \kokkos and \parthenon submodules),
i.e., using a two-stage, second-order method consisting of RK2 integration, piecewise linear
reconstruction and HLLE Riemann solver.

\subsection{Block and communication buffer packing}
\label{res:pack-in-one}
\begin{figure}
\centering
\includegraphics[width=0.5\textwidth]{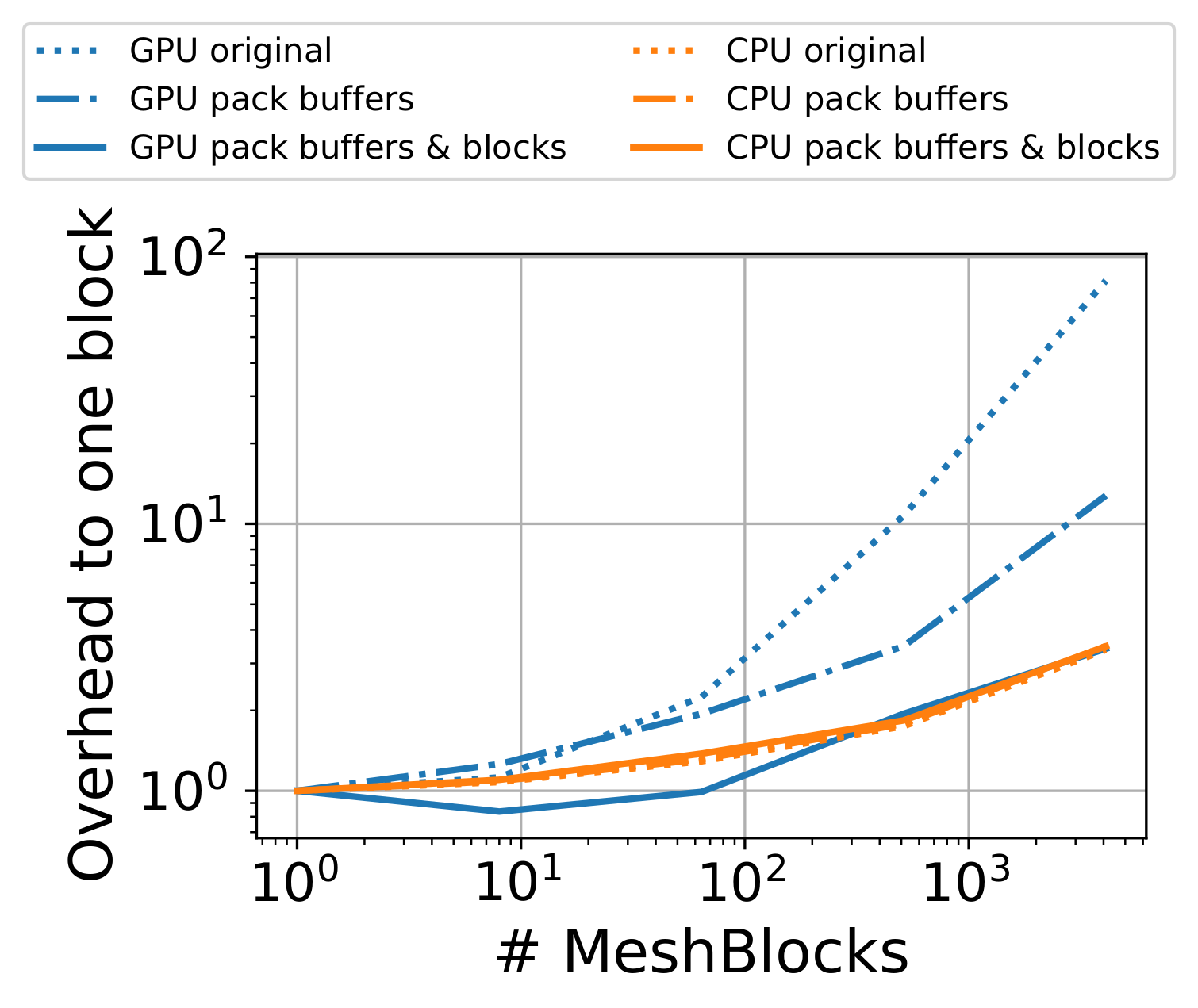}
\caption{Overhead associated with an overdecompositon of the mesh
  measured as relative performance to a second-order MHD update with
  \athenapk using a single meshblock for the entire mesh.
  The mesh size is fixed to $256^3$ ($128^3$) and the block size
  varies from $256^3$ ($128^3$) to $16^3$ ($8^3$) using a single
  process on a single V100 GPU (single Xeon Gold 6148 CPU core).
  The dotted lines show the original performance using a single
  kernel per block and buffer.
  The dash-dotted lines show the performance packing all communication
  buffers of a meshblock in a single kernel and the solid lines
  correspond to using a single kernel to pack all buffers of all
  meshblocks in a single kernel.
  {Performance on CPUs is effectively independent of buffer and block
  packing (all CPU lines are on top of each other).}
}
\label{fig:overdecomp}
\end{figure}

To highlight the need for packing meshblocks and combining the 
communication buffer filling routines in order to improve the 
performance on GPUs, we measured the overhead  associated with
an overdecompositon of the mesh.
In this idealized setup, the mesh size is kept fixed and the meshblock size
is varied.
With smaller and smaller meshblocks, the ratio of ghost cells to active cells
increases, the number of buffers increases, and, generally, the overhead associated
with managing the entire hierarchy of meshblocks increases.
Fig.~\ref{fig:overdecomp} illustrates the relative performance on a single GPU (V100)
and a single CPU core (Xeon Gold 6148)  when going from using a single meshblock
for the entire mesh to 4{,}096 meshblocks.

On the CPU this overdecomposition results in an overhead of $\approx3.5\times$
independent of whether no packing (``original''), packing all buffers of a meshblock
in a single kernel, or packing all buffers of all meshblocks in a single kernel is used.
This is comparable to the original implementation in \athena with an overhead of
$\approx3.3\times$, cf., Fig.~36 in \citet{athenapp}.

On the GPU, the original implementation that launched one kernel per buffer results
in a significant overhead and the performance drops by a factor of $\approx 82\times$.
This can be attributed to the kernel launch overhead ($\approx$5-7\,$\mu$s on Summit) that
is longer than the kernel runtime itself -- especially when the communication buffers 
are small, e.g., for small meshblock sizes or for corners (8 cells) in general.
To alleviate this bottleneck, we first tried to use multiple streams and launching kernels
from multiple threads.
While the performance improved with multiple kernels running simultaneously, the results
were not satisfactory because the kernel launch itself is inherently serial at the CUDA 
level.
The seconds approach of reducing the number of kernel launches by filling all buffers
of a meshblock in one kernel (see Sec.~\ref{sec:boundaries}) and by packing multiple blocks
(see Sec.~\ref{sec:packing}) significantly reduced the additional overhead.
As shown in Fig.~\ref{fig:overdecomp}, filling buffers in a single kernel reduced the
overhead from $\approx 82\times$ to $\approx 13\times$ at an overdecomposition of 4{,}096
blocks.
Combining this with also handling all meshblocks in a single kernel reduced the overhead
further down to $\approx3.5\times$, which is now on par with the CPU result.

\subsection{Pack sizes and overdecomposition}
\label{sec:res-pack-sizes}
\begin{table*}[!t]
\small \sf \centering
\caption{
  Performance of \phydro in $10^8$ zone-cycles/s/node on 16 Summit nodes
  for fixed mesh sizes and various options to distribute the workload.
  The uniform mesh size is fixed to 2{,}048x1{,}536x1{,}024 (1{,}792x384x256) on GPUs (CPUs)
  split into blocks of 
  256$^2$x512, 256$^{3}$, and 128$^3$ per GPU
  (64$^3$, 64$^2$x32, and 32$^2$x16 per CPU core) for 
  1, 2 and 16 blocks per device, respectively.
  The multilevel mesh is identical on GPUs and CPUs and contains a
  cubic region with side length 0.4 refined to level 3 in a unit cube.
  The root grid has a resolution of 256$^3$ and the block size is 32$^3$.
  The resulting mesh hierarchy has 296, 1216, 1{,}352 and 21{,}952 blocks on level
  0, 1, 2, and 3, respectively.
  The ``B'' for the number of \texttt{MeshBlockPack}s per rank stands for using
  one \texttt{MeshBlockPack} for each \texttt{MeshBlock}.
  Using more than one rank per GPU on Summit is enabled by the NVIDIA Multi-Process Service (MPS).
}
\label{tab:distribute-work}
\begin{tabular}{l c c c c c c c c c c c} 
\toprule
 & \multicolumn{7}{c}{Uniform mesh} & \multicolumn{4}{c}{Multilevel mesh} \\
\cmidrule(lr){2-8} \cmidrule(rl){9-12} 
\textbf{\# blocks per dev.} & 1 & \multicolumn{2}{c}{2} & \multicolumn{4}{c}{16} & \multicolumn{4}{c}{259 (GPU) \& 37 (CPU)}\\
\cmidrule(lr){2-2} \cmidrule(rl){3-4} \cmidrule(rl){5-8} \cmidrule(rl){9-12}
\textbf{\# packs per rank} & 1 & 1 & B & 1 & 2 & 4 & B & 1 & 2 & 4 & B\\
\midrule
1 rank per GPU & 10.8 & 11.7 & 10.7 & 11.7 & 11.3 & 11.0 & 9.1 & 2.2 & 2.2 & 2.2 & 1.0 \\
2 ranks per GPU & -- & 12.9 & -- & 12.6 & 12.6 & 12.2 & 11.6 & 2.9 & 3.0 & 3.0 & 1.7 \\
4 ranks per GPU & -- & -- & -- & 13.0 & 13.1 & 12.9 & 12.9 & 3.9 & 4.0 & 4.0 & 2.7 \\
1 rank per CPU core & 0.45 & 0.47 & 0.44 & 0.25 & 0.29 & 0.29 & 0.29 & 0.42 & 0.43 & 0.42 & 0.40 \\
\bottomrule
\end{tabular}
\end{table*}

As already noted in the \athena method paper \citep{athenapp}, some (limited amount of)
overdecomposition, i.e., using more than one block per computing device (e.g., a CPU core) 
resulted in higher performance as it allowed for additional overlapping of computation 
and communication.
However, with an increasing number of blocks per device the block size itself decreases
resulting is a smaller ratio of active to ghost cells that need to be communicated.
Thus, an optimal mesh decomposition is problem and hardware dependent.

For \parthenon with support for running on GPUs and packing multiple blocks into
a \texttt{MeshBlockPack} that are handled simultaneously, finding an optimal decomposition
is even more complex.
This is illustrated in Table~\ref{tab:distribute-work} where we list the
performance per node of \phydro for uniform and multilevel mesh runs on 16 Summit nodes
for various options to distribute the workload.
Note, the example mesh and block sizes are chosen to illustrate a general direction
and details will vary with other factors including (but not limited to) 
devices (and their features), interconnects, mesh hierarchy or block sizes.

When using a single \mpi rank per GPU the best performance is typically
achieved when using just a single pack on each device containing all blocks.
Moreover, in the uniform mesh case overdecomposing the mesh into 2 blocks per device
increases performance from $10.8\times10^8$ zone-cycles/s/node to $11.7\times10^8$.
This also holds for using 16 blocks per device on GPUs as the ratio of
active to ghost cells is still large for block size of 128$^3$.
In contrast, using 16 blocks per CPU core significantly reduces this ratio as
the block size is reduced to 32$^2$x16 in the example given and the
performance drops by $\approx$50\% compared to using 2 blocks per CPU core,
which is optimal (and similar to \athena).

On GPUs the performance can be improved even further when using more than one rank
per device.
However, this needs to be supported by the GPU driver or software as the
\kokkos programming model currently supports a single device per process only.
Both for the uniform and the multilevel mesh the performance is highest when
using 4 ranks per device and splitting all blocks on each rank into two packs.
On the uniform mesh it peaks at $13.1\times10^8$ zone-cycles/s/node and 
for the multilevel mesh at $4.0\times10^8$.
In contrast, the performance for the multilevel mesh is 4$\times$ lower when
using a single rank per GPU handling 256 blocks each and using a separate pack
for each block.
In other words, both packing (i.e., reducing the number of kernel calls) and
using more ranks per device (i.e., reducing the number of blocks per rank
and, in turn, the block management overhead per host rank) each result
in a performance increase of about 2$\times$ in this scenario.
These potential performance gains/losses related to runtime parameters
should encourage problem and application specific tuning for an optimal
use of available computational resources.

\subsection{On-node performance portability}
\label{sec:res-perfport}

\begin{table}[!t]
\small \sf \centering
\caption{
  Performance of \phydro in $10^8$ zone-cycles/s using the full device
  (i.e., either a single GPU or all CPU cores of a node)
  for a typical workload per device on a uniform mesh.
}
\label{tab:perf-port}
\begin{tabular}{l c } 
\toprule
Device (Arch./Instr.) & Performance  \\
\midrule
{AMD MI250X GPU (ROCm, 2x GCD)} & {5.7}\\
NVIDIA A100 GPU (CUDA Cap. 8.0) & 4.2 \\
NVIDIA V100 GPU (CUDA Cap. 7.0) & 2.7  \\ 
AMD MI100 GPU (ROCm) & 2.15  \\
AMD EPYC 7H12 (2x64C x86 AVX2) & 1.45  \\
Intel Xeon 6148 (2x20C x86 AVX512) & 0.67  \\
{IBM Power9 (2x21C)} & {0.51} \\
Intel Xeon E5-2680v4 (2x14C x86 AVX2) & 0.43 \\
Fujitsu  A64FX (1x48C ARMv8.2-A) & 0.36  \\
\bottomrule
\end{tabular}
\end{table}

To highlight the performance portability enabled at the higher level by the
intermediate abstraction layer in \parthenon and at the lower level  by \kokkos,
we measured the performance of \phydro on individual devices across
several architectures.
These include x86 CPUs with AVX2 and AVX512 instruction sets,
ARM CPUs with A64FX architecture,
NVIDIA GPUs and AMD GPUs.

The results are shown in Tab.~\ref{tab:perf-port}.
A single V100 GPU is about $4\times$ faster than a 40 core Intel Skylake system
or $\approx6\times$ faster than a 28 core Intel Broadwell system, which matches
the ratios measured for \kathena \citep{kathena}.
Similarly, the Intel CPU performance of \phydro only about $20\%$ lower than
reported for the same algorithms in \athena \citep{athenapp}
highlighting the low overhead of the abstractions provided by \parthenon.

{On a single MI250X GPU (using 2 GCDs) \phydro is about 2.6$\times$ faster
than on a MI100 GPU and}
on an A100 GPU \phydro is about 55\% faster than on a V100 GPU. This corresponds
to the increased memory bandwidth in combination with the bandwidth limited 
algorithms implemented, cf., the roofline model shown in \citet{kathena}.
While the relative performance of the MI100 GPU with $\approx 80\%$ of a V100 GPU is
still reasonable (despite the $57\%$ increase in memory bandwidth),
the A64FX CPU ($\approx 13\%$ of a V100) is slower than expected based on the
device memory bandwidth.
First tests indicate that some fraction of the lower performance
can be attributed to difficulties of the compiler to (auto)vectorize the compute
kernels, which is in agreement with similar results reported for the
\flash code \citep{Feldman2022}, and, thus, not intrinsic to
the \parthenon framework itself.

\subsection{Scaling results}

All scaling tests in this subsection have been performed with \phydro.
Given the simplicity of the algorithms in the miniapp, \phydro is a well-suited proxy
to gauge the performance of the \parthenon framework itself.
Table~\ref{tab:compiler} lists an overview of the node configuration, software environment,
and compiler flags of all machines used for testing.

Note that the individual mesh sizes used in the scaling tests on uniform meshes
vary slightly between different machines and devices.
We tried to keep the comparison as fair as possible by ensuring that the computational
load per compute element is uniformly distributed, e.g., each compute element (a CPU core
or a GPU) handles the same number of \lstinline!MeshBlock!s for a given test case
so that there is no artificial load imbalance.
{Finally, the numbers reported correspond to the median performance of several tens
  of cycles to mitigate external effects (such as network congestion) as
most of data was collected using non-exclusive allocations.}

\begin{table*}[!t]
\small \sf \centering
\caption{Summary of hardware configuration, software environment and compiler flags used in scaling tests.
  Summit and {Frontier} are operated by the Oak Ridge Leadership Computing Facility,
Booster refers to the JUWELS Booster module operated by the Jülich Supercomputing Centre,
{Frontera is operated by the Texas Advanced Computing Center,}
and Ookami is hosted by the Institute for Advanced Computational Science at Stony Brook University.
}
\label{tab:compiler}
\begin{tabular}{l p{3.5cm} p{2.7cm} p{7cm}}
\toprule
\bfseries Machine &  \bfseries Node conf. & \bfseries Environment & \bfseries Compiler optimization flags \\
\midrule
Summit GPU & \multirow{2}{3.5cm}{2x 22-core Power9 CPU, 6x V100 16GB, NVLink,  2x EDR InfiniBand}  &\multirow{2}{3cm}{GCC 9.1.0, CUDA~11.0.3, SpectrumMPI 10.4.0.3} & 
\texttt{-O3 -mcpu=power9 -mtune=power9 -expt-extended-lambda -Wext-lambda-captures-this -arch=sm\_70}
\\
Summit CPU & &  &
\texttt{-O3 -mcpu=power9 -mtune=power9 -fopenmp-simd -fprefetch-loop-arrays}
 \\
 \\
 Booster GPU & \multirow{2}{3.5cm}{2x 24-core Epyc 7402 CPU, 6x A100 40GB, NVLink3,   4x HDR200 InfiniBand}  &\multirow{2}{3cm}{GCC 11.2.0, CUDA~11.5, OpenMPI~4.1.1} & 
\texttt{-O3 -march=znver2 -mtune=znver2 -expt-extended-lambda -Wext-lambda-captures-this -arch=sm\_80}
\\
Booster CPU & &  &
\texttt{-O3 -march=znver2 -mtune=znver2 -fopenmp-simd -fprefetch-loop-arrays}
 \\
 \\
 {Frontier GPU} & \multirow{1}{3.5cm}{{1x 64-core 3rd Gen EPY, 4x MI250X, Infinity Fabric (xGMI), Slingshot-11}}   &  \multirow{1}{3cm}{{HIP~5.1.20531, ROCm~5.1.0, Cray~MPICH~8.1.17}}  & \texttt{{-O3 -march=znver2 -mtune=znver2 -fno-gpu-rdc --amdgpu-target=gfx90a}}
 \\
 \\
 \\
 {Frontera} & \multirow{1}{3.5cm}{{2x 28-core Intel Xeon Platinum 8280, 1x HDR100 InfiniBand}} & \multirow{1}{3cm}{{ICC~19.1.1.217, Intel~MPI~19.0.9}}  & \texttt{{-O3 -xCORE-AVX512 -qopenmp-simd -qoverride-limits}} 
 \\
 \\
 \\
 Ookami & 1x 48-core Fujitsu A64FX, 1x HDR200 InfiniBand & Fujitsu FCC 4.5.0,  OpenMPI 4.0.1  & \texttt{-Nclang -O3 -ffj-fast-matmul -ffast-math -ffp-contract=fast -ffj-fp-relaxed -ffj-ilfunc -fbuiltin -fomit-frame-pointer -finline-functions -ffj-preex -ffj-zfill -ffj-swp -fopenmp-simd} 
 \\
\bottomrule
\end{tabular}
\end{table*}

\subsubsection{Weak scaling on uniform grids}
\begin{figure}
\centering
\includegraphics[width=\columnwidth]{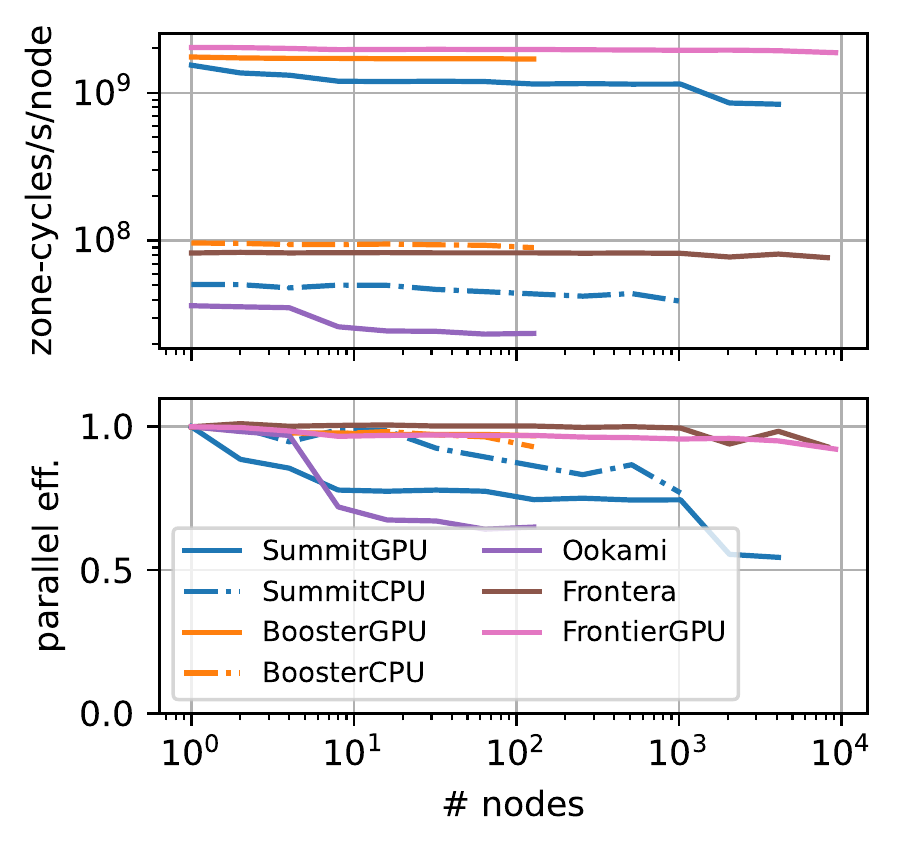}
\caption{Weak scaling of \phydro on uniform grids on various supercomputers with
raw performance in zone-cycles per second per node (top), parallel
efficiency (bottom).
On Summit GPUs (CPUs) on each node handled approximately 586$^3$ (222$^3$) cells,
on JUWELS Booster 812$^3$ (233$^3$),
on Ookami 233$^3$, {on Frontera 245$^3$, and on Frontier 1{,}024$^3$,} respectively.
}
\label{fig:scaling-weak-uni}
\end{figure}
The weak scaling of \phydro on various machines is  illustrated in Fig.~\ref{fig:scaling-weak-uni}.
In general, we used problem sizes that used a large fraction of the available
GPU memory (512x256$^2$ per 16G V100 GPU, 512$^2$x256 per 40G A100, {and 512$^3$ per 64G MI250X GCD})
and 64$^3$ per CPU core.
At the largest scale, \phydro reaches a {92\% parallel efficiency going from one to
9{,}216 nodes (73{,}728 logical GPUs) on Frontier for a total of $1.7\times10^{13}$ zone-cycles/s} --
in other words, effectively updating a {$16{,}384^3$ mesh about four times} per second.
{At the largest rank count, \phydro reaches a 93\% parallel efficiency going from one
to 8{,}192 nodes (458{,}752 MPI ranks with one rank per core) on Frontera.}
Overall, we see a significant speedups using GPUs over CPUs even at large node counts, e.g.,
$\approx29\times$ on a 1{,}024 Summit nodes.
In addition, the parallel efficiency is generally {comparable between CPUs and GPUs with
the exception of Summit.}
This is in agreement with the scaling behavior of \kathena \citep{kathena} and
can be attributed to {the improved node design of more recent machines.
On Frontier and JUWELS Booster each GPU is directly connected to a separate interconnect
card whereas on Summit six GPUs share two InfiniBand cards per node connected to the CPU.
}

\subsubsection{Strong scaling on uniform grid}
\begin{figure}
\centering
\includegraphics[width=\columnwidth]{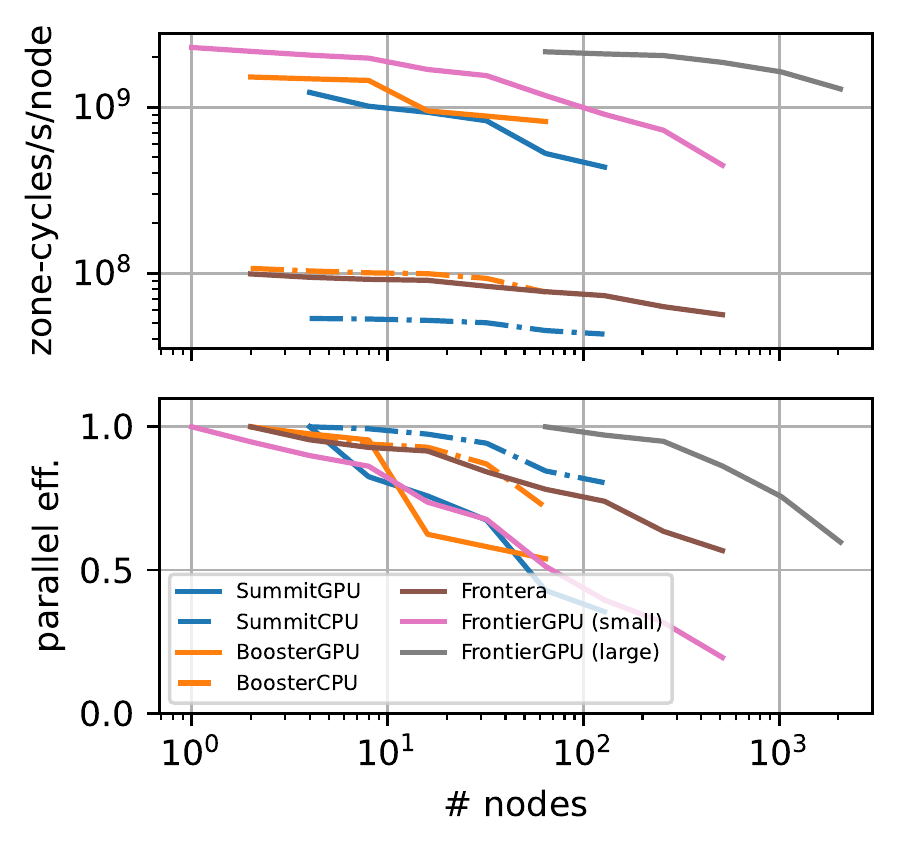}
\caption{Strong scaling  of \phydro on uniform grids on various supercomputers with
raw performance in zone-cycles per second per node (top), parallel
efficiency (bottom).
On Summit GPUs (CPUs) the mesh size was fixed to 1{,}024$^2$x768 (1{,}024x896x768) 
and the load per node varied from 586$^3$ to 185$^3$ (561$^3$ to 177$^3$). 
On JUWELS Booster GPUs (CPUs) the mesh size was fixed to 1{,}024$^3$ (1{,}024$^2$x768) 
and the load per node varied from 813$^3$ to 256$^3$ (738$^3$ to 236$^3$). 
{On Frontera the mesh size was fixed to 1{,}024$^2$x896 and the load per node varied from
  777$^3$ to 122$^3$.
  On Frontier the small (large) mesh size was fixed to 1{,}024$^3$ (4{,}096$^3$) 
and the load per node varied from 1{,}024$^3$ to 128$^3$ (1{,}024$^3$ to 322$^3$).
}
}
\label{fig:scaling-strong-uni}
\end{figure}
The strong scaling on uniform grids of \phydro on various machines is 
illustrated in Fig.~\ref{fig:scaling-strong-uni}.
We used comparable problem sizes of~$\lesssim 1{,}024^3$ and started with
the minimum number of nodes required on each machine.
In general, the parallel efficiency using CPUs is slightly higher than using GPUs on the same
machine, e.g., on Summit remaining at $\approx80\%$ on CPUs for a $32\times$ increase in node
count (going from 4 to 128 nodes).
While the parallel efficiency on Summit drops to $\approx 35\%$ at 128 nodes using GPUs
the raw performance of the GPU accelerated simulations is still more than $10\times$
greater than using CPUs.
The differences between CPU and GPU strong scaling parallel efficiency 
can be attributed to the significantly larger ratio of throughput and memory bandwidth
to problem size on GPUs resulting in a more challenging baseline on GPUs, cf., similar
results for \kathena \citep{kathena}.
{Nevertheless, for a $32\times$ increase in node count on Frontier the parallel
efficiency remains high at 67\% and 60\% going from 1 to 32 or 64 to 2{,}048 nodes, respectively.}

\subsubsection{Strong scaling with multilevel grids}
\begin{figure}
\centering
\includegraphics[width=\columnwidth]{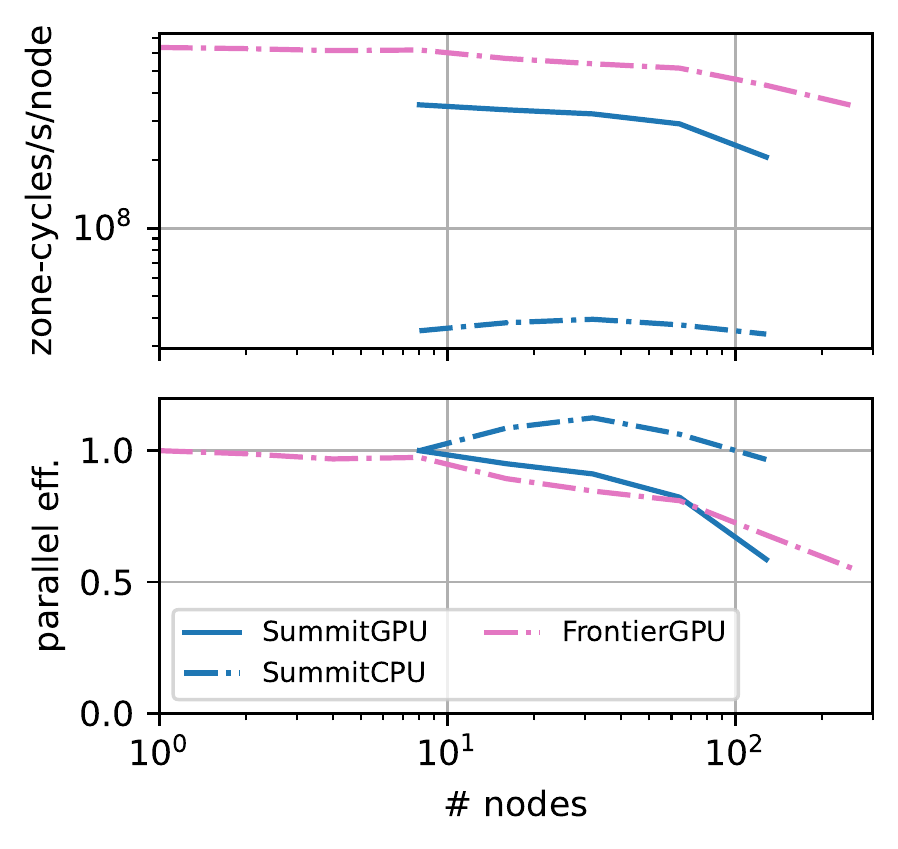}
\caption{Strong scaling of \phydro on multilevel grids on Summit with
raw performance in zone-cycles per second per node (top), parallel
efficiency (bottom).
The mesh is identical to the one presented in Sec.~\ref{sec:res-pack-sizes}, i.e.,
a $256^3$ root grid with $32^3$ blocks and 3 additional levels of refinement resulting in
296, 1216, 1{,}352, and 21{,}952 blocks on level 0, 1, 2, and 3, respectively.
}
\label{fig:scaling-strong-ml}
\end{figure}
In contrast to the scaling tests on uniform grids presented in the previous two 
subsections, Fig.~\ref{fig:scaling-strong-ml} show the strong scaling behavior
of \phydro for a multilevel grid.
The grid is the same as used in Sec.~\ref{sec:res-pack-sizes}, i.e., a block size
of $32^3$ is used on a $256^3$ root grid with 3 additional levels of refinement resulting
in 296, 1216, 1352, and 21952 blocks on level 0, 1, 2, and 3, respectively.
Therefore, this tests now includes the prolongation/restriction machinery for ghost
zones across level boundaries as well as flux correction for faces across level
boundaries.

The strong scaling parallel efficiency on CPUs is generally better than on GPUs on Summit
reaching $\approx97\%$ and $\approx59\%$, respectively, going from 8  to 128 nodes.
Again, simulations on GPUs are significantly faster than ones using CPUs, but the speedup
is lower than on uniform grids, e.g., $\approx10\times$ on 8 nodes and $\approx6\times$
on 128 nodes for the given setup.
This difference stems from the small kernels sizes, e.g., in the flux correction step,
which currently is still follows a ``one kernel per face'' approach, and the
associated overhead.
We expect further improvements by also using the packing approach described in 
Sec~\ref{sec:packing} for the flux correction.
{The (limited) super-linear speedup observed in the CPU runs on Summit can be
attributed to the mesh management overhead where at the smallest scales (8 nodes)
each rank handles $\approx$ 74 blocks, which is successively reduced with larger rank
count, cf., Sec.~\ref{res:pack-in-one} and \ref{sec:res-pack-sizes}.
On GPUs this is not observed as the overhead is hidden by asynchronously running kernels
over larger packs of blocks.
Finally, on Frontier a 256$\times$ increase in resources still results in a parallel
efficiency of 55\% again highlighting the importance of the direct connection between
interconnect and GPUs.
}

\section{Software engineering}
\label{sec:sweng}
\subsection{Development model}
\parthenon is an open, community-driven effort to create a performance-portable
AMR framework applicable for a wide variety of applications. Developers
come from several institutions, have access to different
computational resources, and have different application needs.  In order
to meet the needs of disparate interests within the community, we
enforce sustainable collaborative software practices. 
These are also documented in the repository itself in the development guide.

Collaborative development is facilitated via the \parthenon repository on \github\footnote{
\url{https://github.com/parthenon-hpc-lab/parthenon}}. Each
contribution to the developmental branch is verified with a continuous
integration pipeline, and reviewed and approved by developers from multiple downstream
applications. A consistent code style is strictly
enforced across the code base with each contribution using automated
code style checking and formatting.

New features to the AMR framework are documented and demonstrated in examples
contained within the repository. These examples are then used in continuous
integration testing.

\subsection{Testing}

At the highest level \parthenon uses a \ctest based testing environment
that handles various test cases.
A shorter test suite is triggered automatically for new commits and/or
opened pull request.
An extended test suite is executed nightly for the \verb=develop= branch
or be triggered manually.
Similarly, the test suite can also be triggered locally during development
and offers flexible options to adapt to local environments, e.g.,
with respect to the number of GPUs per node or the \mpi launch command.
The test infrastructure contains the following three building blocks.

\subsubsection{Simple, standard tests} include unit testing,
build testing, and coding style.

For each new feature, developers are encouraged to provide separate unit
tests that are ideally independent of other components in \parthenon.
\parthenon uses \catch for these tests to automatically create descriptive
test cases that integrate with \ctest.

Given the various hardware architectures \parthenon targets and their respective
recommended compilers, the automated build testing covers several
combinations.
These include \verb=Release= and \verb=Debug= builds for NVIDIA GPUs with \nvcc,
x86 CPUs with \gcc, and AMD GPUs with \hipcc.
The builds are tested in \docker containers that are maintained and published
through the main repository so that they are easily available for developers
and users.

Finally, consistent code style is automatically enforced using \clangformat and \cpplint.

\subsubsection{Regression tests} also include integration tests as they
cover more complex use cases.
The majority of regression tests use the examples available in \parthenon to
verify correctness either against the analytic/exact solution or against a good
known previous reference solution.

In contrast to the simple tests, which are directly called from \ctest, we development
a \python based framework for the regression tests.
This framework allows to create complex tests that are tailored to \parthenon, e.g., with
respect to calling a \parthenon based executable (i.e., one of the examples) with a given
input file.
The latter can also be modified from within the testing framework.
Moreover, the ``analysis'' step of each test case also allows to process the test results
(including the data written to disk or the terminal output) to create artifacts
for easy visual inspection.

The testing framework is fully documented and can easily be reused in downstream codes.
This allows for a seamless integration of \parthenon and downstream code testing with
a unified approach.

\subsubsection{Performance testing} and reporting is also realized through a
separate framework: the \parthenon Performance Metrics App (PPMA).
It is a custom \github application whose source is maintained in the main repository.
It allows to run multi-node performance regression tests on internal machines and
can only be triggered manually manually after code review for security reasons.
For each run \json file is created containing information about time and date of the test,
the branch, the commit hash, and various performance metrics.
The results are automatically compared and plotted against the previous five commits of that
branch and against develop.

\section{Current limitations and future enhancements}
\label{sec:limitations}
In the active, ongoing development of \parthenon we already identified
several areas and features that can be further improved and/or need to be implemented
motivated by a downstream code requirement.

For example, \parthenon itself currently only supports Cartesian coordinate systems with
fixed mesh spacing.
Nevertheless, all coordinate related functions are already abstracted and contained in
in a separate class.
Similarly, all functions provided by \parthenon are already making use of those abstraction, e.g.,
when calculating the divergence of a flux or during flux correction in simulations with
mesh refinement.
Therefore, the addition of other coordinate systems is straightforward.

Similarly, the \lstinline!Variables! class is already prepared to handle additional
variable types such as face centered or edge centered variables.
While basic support for face variables is already implemented (covering allocation and
index handling) the boundary and communication routines are not fully refactored yet.

From a performance point of view, we are currently evaluating further improvements in
the ghost zone communication routines.
For example, while overlapping computation and communication is already supported through
the tasking infrastructure in combination with asynchronous MPI routines, all
all ghost zones are currently handled in the same way.
This is not ideal as ghost zones with neighbors on the same rank, i.e., ones that are
directly copied to the receiving buffer, are handled in the same kernels as those
who are first copied into a buffer in preparation for being sent via MPI.
We expect that a split of the kernel into handling remote ghost zones and rank-local
ones separately (in that order) to be more efficient because rank-local buffers would
be copied while all remote buffers are already being transmitted.
The same pattern also applies to the unpacking of the receiving buffers in reverse order.

Independently, first tests indicate that the optimal loop pattern for these buffer handling
kernels depend on many factors including overall simulation setup (e.g.,
\lstinline!MeshBlock! size or  ghost zone width), implementation details (e.g., 
number of components in a \lstinline!Variable! vector), or device architecture.
The results are not yet conclusive, but we expect to eventually provide both 
an architecture specific default pattern as well as a general simulation/algorithm
dependent guideline.
This similarly applies to other runtime parameters such as the default
\lstinline!MeshBlockPack! size or the number of ranks per device,
cf.,~Sec.~\ref{sec:res-pack-sizes}.

\section{Conclusions}
\label{sec:conclusions}
In this article, we presented the performance portable block-structured
adaptive mesh refinement framework \parthenon.
Performance portability is achieved through the use of the \kokkos library
in combination with an intermediate abstraction layer.
The mesh refinement machinery is based on \athena.

The overall design philosophy follows a device-resident approach, i.e., all
simulation data is only allocated on the computing device to reduce data movement.
Moreover, \parthenon is designed for shared capabilities between various downstream
application codes by exposing granular interfaces to the application developers.
At the same time, we strive to keep \parthenon simple enough to be  easily extensible.

Key features includes abstractions for packages, which can be considered as disparate
components containing, for example, a hydrodynamics solver or a radiation transport solver,
abstractions for multidimensional variables including vectors and tensors with
support for sparse allocation, and
a task based applications driver with support for asynchronous, dependency-based task
execution.

From a performance point of view, the key features include the packing of variables
and blocks into larger logical structures so that they can be handled within a single
kernel.
This is particularly relevant for kernels pertaining to filling communication buffers and
when using small block sizes as the number of individual kernel launches can be 
significantly reduced.
Similarly, asynchronous, one-sided MPI communicators are used directly from buffers in device
memory to allow for an overlap of compute kernels and data transfer between nodes.

We demonstrated the success of these design decisions and features in various scaling test
using the hydrodynamics miniapp \phydro reaching a total of {$1.7\times10^{13}$ zone-cycles/s
on 9{,}216 Frontier nodes (73{,}728 logical GPUs) at $\approx92\%$} weak scaling parallel efficiency
(starting from a single node).
Moreover, we demonstrated performance portability across different CPU and GPU
architectures including AMD and NVIDIA GPUs, Intel and AMD x86 CPUs, IBM Power9 CPUs,
and Fujitsu A64FX CPUs.
In general, simulations employing GPUs are significantly faster compared to using the
CPU resources on the same node for both uniform grids (and weak and strong scaling)
as well as for multilevel grids, i.e., setups that require prolongation/restriction
and flux correction.

Several downstream applications are in active development ranging from compressible
magnetohydrodynamics to general relativistic neutrino radiation magnetohydrodynamics
to multi-material compressible hydrodynamics exemplifying the diverse application
scenarios enabled by \parthenon.
In addition, we also introduced the \phydro miniapp supporting full 3D compressible
hydrodynamics with adaptive mesh refinement using \parthenon's capabilities in
just over 1000 lines of code.
This also highlights the use of \parthenon as basis for rapid prototyping and
testing of new algorithms.

Finally, we emphasize that \parthenon is an open, collaborative project and that
new members/contributions are always welcome!

\begin{acks}
The authors would like to thank the \athena team, in particular Kengo Tomida, Kyle Felker, and Chris White
for having provided an open, well-engineered basis for \parthenon.
We also thank the \kokkos team for their continued support throughout the
project {and John Holmen for supporting the scaling tests on Frontier}.
Moreover, we would like to thank Daniel Arndt, Kyle Felker, Max Katz, and Tim Williams
for their contribution to this work during the Argonne GPU Virtual Hackathon 2021.
Finally, we would like to thank our lovely bots, especially \codename{par-hermes}, who is a very good bot.
This work has been assigned a document release number LA-UR-22-21270.
\end{acks}

\begin{dci}
The author(s) declared no potential conﬂicts of interest with respect to the research,
authorship, and/or publication of this article.
\end{dci}

\begin{funding}
This work was supported by the U.S. Department of Energy through the Los Alamos National Laboratory (LANL). LANL is operated by Triad National Security, LLC, for the National Nuclear Security Administration of U.S. Department of Energy (Contract No. 89233218CNA000001).
PG acknowledges funding from LANL through Subcontract No.: \texttt{615487}.
This project has received funding from the European Union's Horizon 2020 
research and innovation programme under the Marie Skłodowska-Curie grant 
agreement No \texttt{101030214}.
{This research used resources of the Oak Ridge Leadership Computing Facility at the Oak Ridge National Laboratory, which is supported by the Office of Science of the U.S. Department of Energy under Contract No. DE-AC05-00OR22725.
  The authors acknowledge the Texas Advanced Computing Center (TACC) at The University of Texas at Austin for providing HPC resources during the April 2022 Texascale Days.
}
Code development, testing, and benchmarking was made possible through various computing
grants including allocations on OLCF Summit and Frontier (\texttt{AST146}), Jülich Supercomputing Centre 
JUWELS (\texttt{athenapk}), Stony Brook's Ookami (\texttt{BrOs091321F}), and
Michgian State University’s High Performance Computing Center.
\end{funding}

\bibliographystyle{SageH}
\bibliography{references.bib}

\end{document}